\documentclass[showpacs,aps,prd,nofootinbib,showkeys,preprintnumbers,superscriptaddress,twocolumn]{revtex4-1}
\usepackage{graphicx}
\usepackage{bm}
\usepackage{amssymb}
\usepackage{amsmath,latexsym}
\usepackage[usenames]{color}
\usepackage{subfigure}
\usepackage{subfigure}
\usepackage{physymb}
\usepackage{slashed}
\usepackage{multirow,array}
\usepackage{mathtools}
\usepackage{mathrsfs}
\usepackage[colorlinks=false,linktocpage=true]{hyperref}
\usepackage{hyperref}
\usepackage[utf8]{inputenc}
\usepackage{lipsum}
\newcommand{\be}{\begin{equation}}
\newcommand{\ee}{\end{equation}}
\newcommand{\bea}{\begin{eqnarray}}
\newcommand{\eea}{\end{eqnarray}}
\newcommand{\beal}{\begin{align}}
\newcommand{\enal}{\end{align}}
\newcommand{\bs}{\begin{subequations}}
\newcommand{\es}{\end{subequations}}
\newcommand{\besp}{\begin{split}}
\newcommand{\eesp}{\end{split}}
\newcommand{\idd}{\indent\indent}
\newcommand{\del}{\partial}

\newcommand{\pxp}{(- p \cdot \Xi \cdot p)}

\newcommand{\dft}{\delta \tilde f}

\newcommand{\up}{u \cdot p}

\newcommand{\mzp}{(- z \cdot p)}

\newcommand{\ene}{\mathcal{E}}

\newcommand{\PL}{\mathcal{P}_L}
\newcommand{\Pperp}{\mathcal{P}_\perp}

\newcommand{\Peq}{\mathcal{P}_{\eq}}
\newcommand{\Pavg}{\bar{\mathcal{P}}}
\newcommand{\pperp}{p^{\{\mu\}}}

\newcommand{\nabperp}{{\nabla_\perp}}
\newcommand{\Wperp}{W^\mu_{\perp z}}
\newcommand{\piperp}{\pi^\munu_{\perp}}
\newcommand{\munu}{{\mu\nu}}

\newcommand{\eq}{\mathrm{eq}}

\newcommand{\um}{u^\mu}
\newcommand{\uum}{u_\mu}
\newcommand{\un}{u^\nu}
\newcommand{\unn}{u_\nu}
\newcommand{\tmn}{T^{\mu\nu}}
\newcommand{\piu}{\pi^{\mu\nu}}

\newcommand{\tem}{T}

\newcommand{\tdf}{\delta \tilde{f}}

\newcommand{\R}{\mathcal{R}}

\newcommand{\I}{\mathcal{I}}

\newcommand{\J}{\mathcal{J}}

\newcommand{\xu}{\Xi^{\mu\nu}}

\newcommand{\she}{\pi}
\newcommand{\Pt}{\mathcal{P}_\perp}
\newcommand{\Pl}{\mathcal{P}_L}
\newcommand{\order}{\mathcal{O}}

\begin{document}
\title{(3+1)-dimensional anisotropic fluid dynamics with a lattice QCD equation of state}
\date{\today}

\author{M.~McNelis}
\affiliation{Department of Physics, The Ohio State University, Columbus, OH 43210, USA}

\author{D.~Bazow}
\affiliation{Department of Physics, The Ohio State University, Columbus, OH 43210, USA}

\author{U.~Heinz}
\affiliation{Department of Physics, The Ohio State University, Columbus, OH 43210, USA}
\affiliation{Theoretical Physics Department, CERN, CH-1211 Gen\`eve 23, Switzerland}
\affiliation{ExtreMe Matter Institute (EMMI), GSI Helmholtzzentrum f\"ur Schwerionenforschung, 
                Planckstrasse 1, D-64291 Darmstadt, Germany}

\preprint{CERN-TH-2018-043}

%%%%%%%%%%%%%%%%%%%%%%%%%%%%%%%%%%%%%%%%%%%%%%%%%%%%%%%%%%
\begin{abstract}
Anisotropic hydrodynamics improves upon standard dissipative fluid dynamics by treating certain large dissipative corrections non-perturbatively. Relativistic heavy-ion collisions feature two such large dissipative effects: (i) Strongly anisotropic expansion generates a large shear stress component which manifests itself in very different longitudinal and transverse pressures, especially at early times. (ii) Critical fluctuations near the quark-hadron phase transition lead to a large bulk viscous pressure on the conversion surface between hydrodynamics and a microscopic hadronic cascade description of the final collision stage. We present a new dissipative hydrodynamic formulation for non-conformal fluids where both of these effects are treated nonperturbatively. The evolution equations are derived from the Boltzmann equation in the 14-moment approximation, using an expansion around an anisotropic leading-order distribution function with two momentum-space deformation parameters, accounting for the longitudinal and transverse pressures. To obtain their evolution we impose generalized Landau matching conditions for the longitudinal and transverse pressures. We describe an approximate anisotropic equation of state that relates the anisotropy parameters with the macroscopic pressures. Residual shear stresses are smaller and are treated perturbatively, as in standard second-order dissipative fluid dynamics. The resulting optimized viscous anisotropic hydrodynamic evolution equations are derived in 3+1 dimensions and tested in a (0+1)-dimensional Bjorken expansion, using a state-of-the-art lattice equation of state. Comparisons with other viscous hydrodynamical frameworks are presented. 
\end{abstract}
%%%%%%%%%%%%%%%%%%%%%%%%%%%%%%%%%%%%%%%%%%%%%%%%%%%%%%%%%%
\pacs{12.38.Mh, 25.75.-q, 24.10.Nz, 52.27.Ny, 51.10.+y}

\keywords{relativistic heavy-ion collisions, quark-gluon plasma, anisotropic hydrodynamics, Boltzmann equation, viscous fluid dynamics}

\maketitle

%%%%%%%%%%%%%%%%%%%%%%%%%%%%%%%%%%%%%%%%%%%%%%%%%%%%%%%%%%
\section{Introduction}
\label{sec1}
%%%%%%%%%%%%%%%%%%%%%%%%%%%%%%%%%%%%%%%%%%%%%%%%%%%%%%%%%%

Dissipative relativistic fluid dynamics has become the workhorse for simulations of the dynamical evolution of relativistic heavy-ion collisions \cite{Heinz:2009xj, Teaney:2009qa, Heinz:2013th, Gale:2013da, Jeon:2015dfa, Bernhard:2016tnd, Alqahtani:2017mhy, Romatschke:2017ejr}. When supplemented with realistic fluctuating initial conditions, a pre-equilibrium evolution module that evolves these initial conditions into starting values for the hydrodynamic evolution, and a hadronic rescattering afterburner that describes the late microscopic kinetic evolution of the collision fireball during its dilute decoupling stage, the approach has yielded impressive quantitative precision in its description of a broad set of soft hadronic observables (i.e. distributions of hadrons with momenta below about $1{-}2.5$\,GeV/$c$) obtained from heavy-ion collision experiments at the Relativistic Heavy Ion Collider (RHIC) and the Large Hadron Collider (LHC) \cite{Muller:2006ee, Nagle:2011uz, Muller:2012zq, Schukraft:2017nbn}, and it has demonstrated convincing predictive power when extending the calculations into new domains of collision energy \cite{Shen:2011eg, Heinz:2011kt, Gale:2012rq, McDonald:2016vlt, Shen:2017bsr} or for new collision systems \cite{Nagle:2013lja, Schenke:2014zha, Habich:2014jna, Romatschke:2015gxa, Shen:2016zpp, Weller:2017tsr}. Surprisingly, the phenomenological success of dissipative fluid dynamics has so far continued unabated in the description of p+Au, d+Au and $^3$He+Au collisions at RHIC and p+p collisions at the LHC \cite{Habich:2014jna, Li:2017qvf, Weller:2017tsr}, i.e. for ``small'' collision systems in which the hydrodynamic model had been widely expected to break down. This finding has generated much recent work addressing two obvious questions arising from these observations: (1) What exactly are the formal criteria that ensure the applicability of relativistic dissipative fluid dynamics to small physical systems undergoing rapid collective expansion and control its eventual break-down? How far away from local thermal equilibrium can a system be and still evolve hydrodynamically? (2) Are there alternate mechanisms at work that can mimic the phenomenological signals of hydrodynamic collective flow, especially in small collision systems, without requiring strong final-state interactions among the constituents of the fireball created in the collision that lead to some degree of approximate local thermalization?

Generically, hydrodynamics is an effective macroscopic theory for the late-time, long-distance evolution of sufficiently equilibrated multiparticle systems. It is typically thought of as a gradient expansion around ideal fluid dynamics. The latter describes locally perfectly thermalized fluids in which any deviation from local thermal equilibrium is immediately erased by strong final-state interactions among the microscopic constituents, i.e. systems whose microscopic relaxation time is effectively zero. Relativistic heavy-ion collisions challenge the validity of such an expansion through extremely large density gradients in the initial state, which lead to explosive collective expansion driving the system away from local thermal equilibrium. Even worse, the ultra-relativistic collision kinematics, combined with the quantum mechanics of the initial particle production process that generates the fireball medium from the energy lost by the colliding nuclei in the collision process \cite{Anishetty:1980zp}, imprints on the system a very strong initial expansion along the ``longitudinal'' beam direction, with an approximately boost-invariant longitudinal expansion velocity profile (``Bjorken flow'' \cite{Bjorken:1982qr}), while any collective expansion in the directions transverse to the beam is initially small and only builds up later in response to transverse pressure gradients. In realistic fluids with non-zero microscopic mean free paths, the resulting large anisotropy in the collective expansion rate causes large anisotropies in the local rest frame (LRF) momentum distributions of the microscopic constituents. In such a situation an expansion around a locally isotropic thermal equilibrium distribution cannot be expected to converge well.

In addition to rapid and strongly anisotropic expansion (which is most problematic during the earliest stage of a heavy-ion collision) another large dissipative effect occurs towards the end of the evolution when the fireball matter undergoes a phase transition from a quark-gluon plasma (QGP) to a hadron resonance gas (HRG). Critical dynamics near the phase transition causes the bulk viscosity to become large and peak near the (pseudo-)critical temperature $T_\mathrm{c}$ \cite{Prakash:1993bt, Paech:2006st, Arnold:2006fz, Kharzeev:2007wb, Karsch:2007jc, Meyer:2007dy,Song:2009rh}. The resulting large bulk viscous pressure is associated with a strong deviation of the LRF momentum distribution from thermal equilibrium. Again, this provides a challenge for any expansion around a local equilibrium distribution function. Since (due to the screening of color interactions by color confinement) the constituents' mean free path in the hadron resonance gas is much larger than in the quark-gluon plasma, the Knudsen number (defined as the product of the mean free time and the scalar expansion rate) increases suddenly as the QGP turns into hadrons, to the extent that the subsequent evolution can no longer be reliably described by hydrodynamics \cite{Song:2010aq}. One must therefore switch to a microscopic kinetic description basically as soon as the hadronization process is complete. At this point the critically enhanced bulk viscous pressure is still large because critical slowing down \cite{Berdnikov:1999ph, Rajagopal:2009yw, Song:2010aq} prohibits it from relaxing quickly to the much smaller values expected away from the phase transition. Its effect on the hadron distribution functions in the HRG can therefore not be treated effectively as a perturbation around local thermal equilibrium and should be accounted for non-perturbatively.

In this work we develop an improved version of anisotropic hydrodynamics that accounts for large shear viscous effects caused by a strong longitudinal-transverse anisotropy of the expansion rate and for large viscous corrections caused by critical dynamics near the quark-hadron phase transition non-perturbatively. The formalism is constructed for full (3+1)-dimensional evolution and tested numerically for (0+1)-dimensional boost-invariant expansion along the beam direction (Bjorken flow \cite{Bjorken:1982qr}). Numerical results for full (3+1)-dimensional evolution of heavy-ion collisions with realistic fluctuating initial conditions \cite{Schenke:2012wb, Gale:2012rq, Liu:2015nwa, Bernhard:2016tnd} will be presented in a future publication. To derive the anisotropic hydrodynamic evolution equations we start from an underlying kinetic theory, the relativistic Boltzmann equation in relaxation time approximation (RTA BE). While such a classical kinetic approach is known to only work for dilute and weakly coupled gases \cite{Arnold:2002zm}, it allows to derive the structure of the macroscopic hydrodynamic equations through a systematic moment expansion \cite{Denicol:2012cn,Bazow:2013ifa,Molnar:2016vvu,Molnar:2016gwq}. As an effective theory, the structure of these equations holds equally well for strongly and weakly coupled systems (i.e. it depends only on the separation of microscopic and macroscopic length scales), as long as one changes the material properties of the fluid (i.e. its equation of state, transport coefficients, relaxation times, etc.) that enter as input into the hydrodynamic description to the actual situation of interest.            
 
Our approach starts from the general treatment described in \cite{Martinez:2010sc, Florkowski:2010cf, Martinez:2012tu, Bazow:2013ifa, Florkowski:2014bba, Molnar:2016vvu}, expanding the Boltzmann equation around an anisotropic local rest frame distribution function $f_a$ which in our case is deformed around the local equilibrium distribution function by {\em two} parameters to account for one large shear stress component and a large bulk viscous pressure as described above. Our main innovation is that, following recent insights reported in \cite{Molnar:2016gwq, Martinez:2017ibh}, the evolution of these deformation parameters is optimized by determining them through generalized dynamical Landau matching conditions, similar to those fixing the evolution of the temperature and chemical potential. This guarantees that the leading order anisotropic distribution $f_a$ (around which the full distribution function is expanded in moments) fully accounts not only for the energy and conserved charge density, but also for the longitudinal and transverse pressures (or, equivalently, the longitudinal-transverse pressure anisotropy (which is the largest shear stress component) and the bulk viscous pressure, as described above). It also significantly simplifies the structure of the relaxation equations for the residual dissipative flows. Writing the full distribution function $f$ as $f=f_a+\tdf$, the deviation $\tdf$ describes the (smaller) residual shear stress component and the charge diffusion current. In the present work we will mostly ignore conserved charges and will hence set the charge chemical potential and charge diffusion effects to zero, leaving a complete treatment to a follow-up paper.

Initially, the resulting hydrodynamic framework is formulated in terms of evolution equations for the parameters characterizing the leading-order distribution $f_a$ (temperature, chemical potential, momentum-deformation parameters), similar to the traditional approach reported in \cite{Martinez:2012tu, Florkowski:2014bba, Alqahtani:2017jwl, Alqahtani:2017tnq} which makes explicit reference to kinetic theory and thus requires that such a distribution function exists and is well-defined. We subsequently break this connection to an underlying kinetic theory by developing a technique that allows to express these microscopic parameters in terms of macroscopic hydrodynamic quantities so that {\em the formalism can be completely formulated and solved as a macroscopic theory}. This procedure introduces the concept of an ``anisotropic equation of state'' (aEOS) which we discuss at some length. We here use a weakly-interacting quasiparticle model with a temperature-dependent particle mass \cite{Alqahtani:2016rth, Tinti:2016bav} to calculate this aEOS and also the required transport coefficients, keeping mind that in later applications to heavy-ion collisions these ingredients should be computed or modeled for QCD, or considered as phenomenological parameters to be determined from the experimental data.

With this work we open the door to answering the question to what extent complete second-order anisotropic fluid dynamics (which we call ``viscous anisotropic hydrodynamics'' or vaHydro \cite{Bazow:2013ifa}) extends the range of validity of dissipative fluid dynamics towards smaller collision systems and earlier switching times between the pre-equilibrium and hydrodynamic stages. Having a non-equilibrium hydrodynamic approach that is optimized to the particular challenges posed by ultra-relativistic collisions between nuclei as small as protons is a necessary step towards developing a quantitatively predictive dynamical framework that can set benchmarks for comparison with experimental data and with other, non-hydrodynamic explanations of the latter. 

Before starting the technical part of our discussion we introduce our notation. Throughout this work we use natural units $\hbar{\,=\,}c{\,=\,}k_B{\,=\,}1$. The metric signature is taken to be ``mostly minus'' $(+,-,-,-)$. The local rest frame (LRF) is defined as the Landau frame: the fluid velocity $u^\mu$ is the normalized time-like eigenvector of the energy-momentum tensor, $T^{\mu\nu}u_\nu{\,=\,}\ene u^\mu$, where $\ene$ is the energy density in the local rest frame. It satisfies  $u^\mu{\,=\,}T^{\mu\nu}u_\nu \big/ \sqrt{u_\mu T^\munu u_\nu}$ and the normalization condition $u_\mu u^\mu{\,=\,}1$. Unless otherwise indicated we ignore conserved charges and their associated chemical potentials.

The paper is structured as follows: In Sec.~\ref{sec2} we briefly review the general structure of anisotropic hydrodynamics and its evolution equations for the energy density and longitudinal and transverse acceleration. Relaxation equations for the anisotropic dissipative flows are derived in Sec.~\ref{sec3}, starting from the relativistic Boltzmann-Vlasov equation in relaxation time approximation and implementing generalized Landau matching conditions for the evolution of the microscopic parameters characterizing the anisotropic distribution function. In Sec.~\ref{sec4} we show how to integrate a realistic lattice QCD equation of state into the anisotropic hydrodynamic framework and reformulate its evolution equations in purely macroscopic form, i.e. without any reference to parameters that are only defined in a kinetic theory model for the microscopic dynamics. In that Section we also discuss the resolution of several technical issues arising in the process of solving the anisotropic hydrodynamic evolution equations numerically. In Sec.~\ref{sec5} we illustrate the performance of our anisotropic hydrodynamic framework in comparison with standard dissipative fluid dynamics for a simple system undergoing Bjorken flow. A summary of our findings and a brief outlook are presented in Sec.~\ref{sec6}. Several appendices supply additional technical ingredients, including (in Appendix~\ref{appd}) the derivation of the evolution equations for standard viscous fluid dynamics needed for the comparison shown in Sec.~\ref{sec5}.

%%%%%%%%%%%%%%%%%%%%%%%%%%%%%%%%%%%%%%%%%%%%%%%%%%%%%%%%%%
\section{Anisotropic Hydrodynamics}
\label{sec2}
%%%%%%%%%%%%%%%%%%%%%%%%%%%%%%%%%%%%%%%%%%%%%%%%%%%%%%%%%%

%%%%%%%%%%%%%%%%%%%%%%%%%%%%%%%%%%%%%%%%%%%%%%%%%%%%%%%%%%
\subsection{Ideal fluid decomposition}
\label{sec2a}
%%%%%%%%%%%%%%%%%%%%%%%%%%%%%%%%%%%%%%%%%%%%%%%%%%%%%%%%%%

In relativistic hydrodynamics, the energy-momentum tensor of a perfect fluid is best decomposed in the basis $u^\mu u^\nu$ and $\Delta^\munu = g^\munu{-}u^\mu u^\nu$ where $u^\mu$ is the fluid four-velocity (i.e. the four-velocity of the local rest frame (LRF) relative to the global frame): 
\be
\label{eq1}
  \tmn(x)=\ene(x)\,\um(x)\un(x) - \Peq(x) \,\Delta^\munu(x).
\ee
From here on we will mostly suppress the space-time ($x$) dependence for notational simplicity. The LRF energy density $\ene$ and thermal equilibrium pressure $\Peq$ are recovered by projecting the energy-momentum tensor onto the temporal and spatial directions: $\ene = \uum\unn\tmn$ and $\Peq = - \frac{1}{3} \Delta_\munu \tmn$. Since the thermal pressure is isotropic none of the spatial directions in the LRF are special, and there is no advantage in decomposing the spatial projector $\Delta^\munu$ further.

%%%%%%%%%%%%%%%%%%%%%%%%%%%%%%%%%%%%%%%%%%%%%%%%%%%%%%%%%%
\subsection{Viscous fluid decomposition}
\label{sec2b}
%%%%%%%%%%%%%%%%%%%%%%%%%%%%%%%%%%%%%%%%%%%%%%%%%%%%%%%%%%

Standard dissipative fluid dynamics is formulated by decomposing $T^\munu$ in the same basis, but adding dissipative corrections accounting for the bulk viscous pressure $\Pi$ and the shear stress tensor $\pi^\munu$:
\be
\label{eq2}
  \tmn=\ene\,\um\un - (\Peq{+}\Pi) \,\Delta^\munu  + \pi^\munu .
\ee
This assumes that the LRF is the Landau frame, i.e. that in the LRF there is no net momentum flow. Eq.~(\ref{eq2}) implicitly assumes that in the LRF all viscous corrections are of similar order of magnitude and small relative to the equilibrium energy density and pressure, such that there still is no advantage in decomposing the locally spatial projector further into individual spatial directions.  

%%%%%%%%%%%%%%%%%%%%%%%%%%%%%%%%%%%%%%%%%%%%%%%%%%%%%%%%%%
\subsection{Anisotropic viscous fluid decomposition}
\label{sec2c}
%%%%%%%%%%%%%%%%%%%%%%%%%%%%%%%%%%%%%%%%%%%%%%%%%%%%%%%%%%

The general arguments presented in the introduction imply that in relativistic heavy-ion collisions the shear stress $\pi^\munu$ is highly anisotropic, and the difference $\PL{-}\Pperp$ between the longitudinal and transverse pressures  can be quite large, due to strongly different longitudinal and transverse expansion rates. This suggests that anisotropic hydrodynamics is best formulated by using a more detailed decomposition of the energy-momentum tensor in which the spatial projector is further decomposed as $\Delta^\munu = \Xi^\munu - z^\mu z^\nu$, where $z^\mu$ points along the beam direction and $\Xi^\munu \equiv g^\munu - u^\mu u^\nu + z^\mu z^\nu$ projects onto the spatially transverse directions in the LRF \cite{Molnar:2016vvu}:
\be
\label{eq3}
  \tmn=\ene\,\um\un + \PL \, z^\mu z^\nu - \Pperp\,\xu  + 2\,W^{(\mu}_{\perp z} z^{\nu)} + \piu_\perp.
\ee
The round parentheses around pairs of Lorentz indices indicate symmetrization: $W^{(\mu}_{\perp z} z^{\nu)} \equiv\frac{1}{2}\bigl(W^{\mu}_{\perp z} z^{\nu}{+}W^{\nu}_{\perp z} z^{\mu}\bigr)$. The anisotropic decomposition (\ref{eq3}) clearly separates the pressures $\PL$ and $\Pperp$ from the other dissipative components. Given an arbitrary energy momentum tensor $T^\munu$, the anisotropic hydrodynamic quantities appearing in this decomposition can be obtained by the following projections:
\bs
\beal
\label{eq4}
  &\ene = \uum\unn\tmn\,, \\
  &\PL =  z_\mu z_\nu\tmn\,,\\
  &\Pperp = - \frac{1}{2}\Xi_{\mu\nu}\tem^{\mu\nu}\,, \\
  &W^{\mu}_{\perp z}\, = - \Xi^{\mu}_{\alpha}\tem^{\alpha\nu}z_\nu\,, \\
  &\piu_\perp= \Xi_{\alpha\beta}^{\mu\nu}\tem^{\alpha\beta} \,,
\end{align}
\es
In the last line we introduced the symmetric traceless transverse projection tensor $\xu_{\alpha\beta} = \frac{1}{2}\big(\Xi^\mu_\alpha\Xi^\nu_\beta+\Xi^\nu_\beta\Xi^\mu_\alpha-\xu\Xi_{\alpha\beta}\big)$. The corresponding transverse shear stress tensor $\piperp$ describes two shear stress degrees of freedom that account for momentum diffusion currents along the transverse directions. It is traceless and orthogonal to both the fluid velocity and the direction of the pressure anisotropy:
\be
\label{eq5}
\begin{aligned}
   \pi^\mu_{\perp,\mu} = u_\mu \piperp = z_\mu \piperp = 0.
\end{aligned}
\ee
Another two shear stress degrees of freedom are encoded in the longitudinal-momentum diffusion current $\Wperp$  which is orthogonal to both $u_\mu$ and $z_\mu$:
\be
\label{eq6}
\begin{aligned}
  u_\mu \Wperp = 0 = z_\mu \Wperp.
\end{aligned}
\ee
The remaining fifth (and largest) shear stress component is given by $\Pl{-}\Pt$. Altogether, the 5 independent components of the standard shear stress tensor in Eq.~(\ref{eq2}) are related to those in the anisotropic decomposition (\ref{eq3}) by
\be
\label{eq7}
  \pi^{\mu\nu}=\she^{\mu\nu}_{\perp} + 2\,W^{(\mu}_{\perp z} z^{\nu)}
  +\frac{1}{3} \big(\Pl - \Pt\big) \big(2z^\mu z^\nu - \xu\big)
\ee
while the single bulk viscous pressure degree of freedom $\Pi$ in Eq.~(\ref{eq2}) is related to the thermal, longitudinal and transverse pressures in Eqs.~(\ref{eq2},\ref{eq3}) by
\be
\label{eq8}
  \Pi = \frac{2\, \Pperp + \PL}{3} - \Peq.
\ee
Here $\Peq$ is not an independent degree of freedom but related to the energy density $\ene$ by the equation of state (EOS) of the fluid, $\Peq(\ene)$. 

%%%%%%%%%%%%%%%%%%%%%%%%%%%%%%%%%%%%%%%%%%%%%%%%%%%%%%%%%%
\subsection{Hydrodynamic evolution equations}
\label{sec2d}
%%%%%%%%%%%%%%%%%%%%%%%%%%%%%%%%%%%%%%%%%%%%%%%%%%%%%%%%%%

Four of the ten evolution equations that control the dynamics of the energy-momentum tensor are obtained from  
the conservation laws for energy and momentum
\be
\label{eq9}
  \del_\mu T^\munu = 0. 
\ee

Projecting with $u_\nu$ on the temporal direction in the LRF provides an evolution equation for the LRF energy density:
\be
\label{eq10}
\begin{split}
  &\dot\ene + (\ene{+}\Pperp) \theta_\perp + (\ene{+}\PL) z_\mu D_z u^\mu \\
  &\ \ + W^{\mu}_{\perp z}\bigl(D_z u_\mu{-}z_\nu \nabla_{\perp\mu}u^\nu\bigr) - \piu_\perp \sigma_{\perp,\munu} = 0.
\end{split}
\ee
Here and below a dot over or a $D$ in front of a quantity denotes the comoving time derivative, e.g. $D\ene\equiv \dot\ene \equiv  u^\mu \del_\mu \ene$. $z_\mu D_z u^\mu$ is the scalar longitudinal expansion rate, and $\theta_\perp{\,=\,}\nabla_{\perp}{\cdot}u$ is the scalar transverse expansion rate. The longitudinal derivative and transverse gradient in the LRF are written as $D_z = - z^\mu \del_\mu$ and $\nabla_{\perp\mu} = \Xi_\mu^\nu \del_\nu$, respectively. $\sigma_{\perp,\munu} = \Xi^{\alpha\beta}_\munu \del_\alpha u_\beta$ is the transverse velocity-shear tensor. 

The longitudinal projection $z_\nu \del_\mu T^\munu = 0$ yields an equation for the longitudinal acceleration of the fluid in the LRF:
\be
\label{eq11}
\begin{split}
(\ene{+}\PL) z_\mu \dot{u}^\mu & =  - D_z \PL + (\PL{-}\Pperp) \tilde\theta_{\perp}\\
  &\quad - W^\mu_{\perp z} D_z z_\mu 
     + \piperp \tilde{\sigma}_{\perp,\mu\nu}.
\end{split}
\ee
Here $\tilde\theta_\perp\equiv \nabperp{\cdot}z$ and $\tilde \sigma_{\perp,\munu} = \Xi^{\alpha\beta}_\munu \del_\alpha z_\beta$.\footnote{%
	 Generically we use tildes to indicate quantities involving derivatives of $z^\mu$ instead of $u^\mu$.
	 }

An equation for the transverse acceleration is obtained from the transverse projection $\Xi^\alpha_\nu \del_\mu T^\munu = 0$:
\be
\label{eq12}
\begin{split}
  & (\ene{+}\Pperp) \Xi^\alpha_\nu \dot{u}^\nu = \nabla_\perp^\alpha \Pperp 
                                                                        + (\PL{-}\Pperp) \Xi^\alpha_\nu D_z z^\nu\\  
  &\ - W^\alpha_{\perp z} \Big(\frac{3}{2} \tilde\theta_\perp{-}z_\nu \dot{u}^\nu\Big) 
      - W_{\perp z, \nu}(\tilde\sigma^{\alpha\nu}_\perp{-}\tilde\omega^{\alpha\nu}_\perp) \\
  &\ + \Xi^\alpha_\nu D_z W^\nu_{\perp z} + \pi^{\alpha\nu}_\perp(\dot{u}_\nu{+}D_z z_\nu)  - \Xi^\alpha_\nu \nabla_{\perp\mu} \piperp .
\end{split}
\ee
Here $\tilde \omega_\perp^{\alpha\nu} \equiv \Xi^{\alpha\mu} \Xi^{\nu\beta} \del_{[\beta} z_{\mu]}$ where the square brackets indicate antisymmetrization: $\del_{[\beta} z_{\mu]} \equiv \frac{1}{2}\bigl(\del_{\beta} z_{\mu}{-}\del_{\mu} z_{\beta}\bigr)$. Equations~(\ref{eq10})-(\ref{eq12}) agree (after adjustment of notation) with Eqs.\,(146)--(148) in Ref.~\cite{Molnar:2016vvu}. 

%%%%%%%%%%%%%%%%%%%%%%%%%%%%%%%%%%%%%%%%%%%%%%%%%%%%%%%%%%
\section{Dissipative relaxation equations}
\label{sec3}
%%%%%%%%%%%%%%%%%%%%%%%%%%%%%%%%%%%%%%%%%%%%%%%%%%%%%%%%%%

To close the system of equations, we need six additional relaxation equations for $\PL$, $\Pperp$, $\Wperp$ and $\piperp$. Their dynamics is not controlled by macroscopic conservation laws but by microscopic interactions among the fluid's constituents. As discussed in Sec.~\ref{sec1}, we will here derive them by assuming a weakly-coupled dilute fluid whose microscopic physics can be described by the relativistic Boltzmann-Vlasov equation for a single particle species with a medium-dependent mass:
\be
\label{eq13}
p^\mu \del_\mu f + m \, \del^\mu m \, \del_\mu^{(p)} f = C[f].
\ee
Here $f(x,p)$ is the single particle distribution function, $C[f]$ is the collision kernel, $m(x)$ is the medium-depen\-dent effective mass, and $\del_\mu^{(p)}$ is the momentum derivative.

%%%%%%%%%%%%%%%%%%%%%%%%%%%%%%%%%%%%%%%%%%%%%%%%%%%%%%%%%%
\subsection{Anisotropic distribution function}
\label{sec3a}
%%%%%%%%%%%%%%%%%%%%%%%%%%%%%%%%%%%%%%%%%%%%%%%%%%%%%%%%%%

For anisotropic hydrodynamics we split the distribution function into a momentum-anisotropic leading-order contribution $f_a(x,p)$ and a small residual correction $\dft$:
\be
\label{eq14}
  f(x,p) = f_a(x,p) + \dft(x,p)
\ee
For the leading order distribution we take the generalized Romatschke-Strickland form \cite{Romatschke:2003ms, Tinti:2013vba, Tinti:2015xwa}:
\be
\label{eq15}
  f_a(x,p) = f_\eq\left(\frac{\sqrt{\Omega_\munu(x) \, p^\mu p^\nu} - \tilde\mu(x)}{\Lambda(x)} \,\right).
\ee
Here $f_{\eq}(z) = 1/(e^z + \Theta)$ is the equilibrium distribution, with $\Theta = 1,0,-1$ for Fermi-Dirac, Boltzmann, and Bose-Einstein statistics, respectively. $\Lambda(x)$ is an effective temperature and $\tilde\mu(x)$ an effective chemical potential. In this work we consider a system without conserved charges and assume, for simplicity, that particle number changing microscopic processes in the collision term $C[f]$ are so fast that the effective chemical potential relaxes to zero faster than any other microscopic time scale.\footnote{%
       This assumption is not realistic: since number-changing processes are a subset of all microscopic processes,
       chemical relaxation processes are typically slower than momentum-changing ones. We here make this
       assumption for simplicity only and intend to relax it in future work.
       }

In this work the leading-order momentum anisotropy is encoded in the ellipsoidal tensor 
\be
\label{eq16}
\begin{split}
   \Omega_\munu(x) = \,\,& u_\mu(x) u_\nu(x) - \xi_\perp(x) \Xi_\munu(x) \\
   & + \xi_L(x) z_\mu (x) z_\nu (x).
\end{split}
\ee
It contains two space-time dependent anisotropy parameters $\xi_L$ and $\xi_\perp$. With $\Xi_\munu p^\mu p^\nu=-p^2_{\perp,\mathrm{LRF}}$ (i.e. the square of the transverse momentum $\bm{p}_{\perp,\mathrm{LRF}}$ in the LRF), $\Omega_\munu p^\mu p^\nu$ can be rewritten as
\be
\label{eq17}
    \Omega_\munu p^\mu p^\nu = m^2 + (1{+}\xi_\perp)p^2_{\perp,\mathrm{LRF}} 
                                                             + (1{+}\xi_L)p^2_{z,\mathrm{LRF}}.
\ee
The difference $\xi_L{-}\xi_\perp$ can be attributed to a manifestation of shear stress (resulting in a difference between the the longitudinal and transverse pressures) while the sum  $\xi_L{+}\xi_\perp$ encodes a bulk viscous pressure \cite{Nopoush:2014pfa,Bazow:2015cha}. Introducing the notation $\alpha_{L,\perp}(x) = \bigl(1+\xi_{L,\perp}(x)\bigr)^{-1/2}$, Eq.~(\ref{eq15}) can be written in LRF momentum components more conveniently as
\be
\label{eq18}
  f_a = f_{\eq}\left(\frac{1}{\Lambda} \sqrt{m^2 + \frac{p^2_{\perp,\mathrm{LRF}}}{\alpha_\perp^2} + \frac{p^2_{z,\mathrm{LRF}}}{\alpha_L^2}}\,\right).
\ee

To make the decomposition (\ref{eq14}) unique one must specify the three parameters $\Lambda(x)$ and $\alpha_{L,\perp}(x)$. We proceed as follows: The physical temperature $T$ of the system is defined by the LRF energy density via the thermodynamic relation $\ene(x) \equiv \ene\bigl(T(x)\bigr)$. To relate the effective temperature $\Lambda$ to $T$ we impose the generalized Landau matching condition\footnote{%
    We note that in the presence of conserved charges the effective chemical potential $\tilde\mu$ would
    be related to the physical chemical potential associated with the conserved charge by a similar Landau
    matching condition for the conserved particle number $n$:
    \be
    \label{eq19}
    %\nonumber
       \delta \tilde n \equiv \langle \up \rangle_{\dft} = 0.
    \ee
    }    
\be
\label{eq20}
  \delta \tilde\ene \equiv \langle (\up)^2 \rangle_{\dft} = 0, 
\ee
which states that  $\Lambda(\bm{\alpha})$ must be chosen such that the residual deviation $\dft$ does not contribute to the energy density. This fixes $\Lambda(\bm{\alpha})$ as a function of $T$; the two agree in the limit $\bm{\alpha}\to1$ when the anisotropic leading-order distribution $f_a$ reduces to a locally isotropic equilibrium distribution $f_\eq(u{\cdot}p/T)$.

The momentum deformation parameters $\alpha_{L,\perp}(x)$ are fixed by similar generalized Landau matching conditions for the transverse and longitudinal pressures
\bs
\label{eq21}
\beal
& \PL = \Peq + \Pi + \pi^{zz}_\mathrm{LRF}, \\
& \Pperp = \Peq + \Pi - {\textstyle\frac{1}{2}} \pi^{zz}_\mathrm{LRF}.
\end{align}
\es 
Here $\pi^{zz}_\mathrm{LRF}$ is the LRF value of the longitudinal diagonal element of the shear stress tensor $\pi^\munu$ in the decomposition (\ref{eq2}). Note that both the bulk viscous pressure $\Pi$ and the shear stress component $\pi^{zz}_\mathrm{LRF}$ are here assumed to be ``large'' such that they must be accounted for already at leading order, by adjusting the parameters $\alpha_{L,\perp}(x)$ accordingly. This is done by demanding
\bs
\label{eq22}
\beal
  & \delta \tilde{\mathcal{P}}_L \equiv \langle \mzp^2 \rangle_{\delta \tilde f} = 0, \\
  & \delta \tilde{\mathcal{P}}_\perp  \equiv  {\textstyle\frac{1}{2}} \langle \pxp \rangle_{\delta \tilde f} = 0.
\end{align}
\es
By imposing these conditions, $\alpha_{L,\perp}(x)$ are adjusted such that the longitudinal and transverse pressures $\PL$ and $\Pperp$ are everywhere fully accounted for by the leading-order distribution $f_a$, with zero residual contributions from $\tdf$. This is an application of the anisotropic matching scheme proposed by Tinti in \cite{Tinti:2015xwa} and a generalization of the $P_L$-matching scheme proposed and studied in \cite{Molnar:2016gwq,Martinez:2017ibh} to both $\PL$ and $\Pperp$ (or, equivalently, to the pressure anisotropy $\PL{-}\Pperp\sim\pi^{zz}_\mathrm{LRF}$ and the bulk viscous pressure $\Pi$). In this matching scheme, the $\dft$ correction generates only the residual dissipative flows described by $\Wperp$ and $\piperp$, which break the cylindrical symmetry of the distribution function in the LRF and account for the remaining four smaller components of the shear stress tensor $\pi^\munu$ in Eq.~(\ref{eq7}).

With the matching conditions (\ref{eq19}), (\ref{eq20}), and (\ref{eq22}a,b) we have the following kinetic theory expressions for the particle number and energy densities as well as for the longitudinal and transverse pressures:\footnote{%
   	The superscript $(k)$ on the thermodynamic quantities indicates their kinetic theory definition for a gas 
	of weakly-interacting quasiparticles. The purpose of this notation will become clear later when we introduce 
	a more realistic EOS.
	}
\bs
\label{eq23}
\beal
  &\label{eq23a} n^{(k)} = \langle \up \rangle_{f_a} = \I_{1000}, \\
  &\label{eq23b} \ene^{(k)} = \langle (\up)^2 \rangle_{f_a} = \I_{2000}, \\
  &\label{eq23c} \PL^{(k)} = \langle \mzp^2 \rangle_{f_a} = \I_{2200}, \\
  &\label{eq23d} \Pperp^{(k)} = {\textstyle\frac{1}{2}} \langle \pxp \rangle_{f_a} = \I_{2010}.
\end{align}
\es
The ``anisotropic integrals'' $\I_{nrqs}$ over the leading-order distribution function $f_a$ that appear in these equations are defined in Eq.~(\ref{eqA1}).

%%%%%%%%%%%%%%%%%%%%%%%%%%%%%%%%%%%%%%%%%%%%%%%%%%%%%%%%%%
\subsection{Relaxation equations I}
\label{sec3b}
%%%%%%%%%%%%%%%%%%%%%%%%%%%%%%%%%%%%%%%%%%%%%%%%%%%%%%%%%%

The relaxation equations for the dissipative flows are obtained by expressing the latter as moments of the distribution function and using the Boltzmann equation to describe its evolution, using the decomposition (\ref{eq14}) and treating $\dft$ as a small perturbation. We start from   
\bs
\label{eq24}
\beal
& \dot{\mathcal{P}}^{(k)}_L = D \int_p \mzp^2 f_a \,, \\
& \dot{\mathcal{P}}^{(k)}_\perp = \frac{1}{2} D \int_p \pxp f_a \,, \\
& \dot{W}^{\{\mu\}}_{\perp z} = \Xi^\mu_\nu D \int_p \mzp \, p^{\{\nu\}} \dft \,, \\
& \dot{\pi}^{\{\munu\}}_{\perp} = \Xi^\munu_{\alpha\beta} D \int_p p^{\{\alpha} p^{\beta\}} \dft \,,
\end{align}
\es
where we defined the compact notations \cite{Molnar:2016vvu}
\begin{eqnarray}
\label{eq25}
  && a^{\{\mu\}} \equiv \Xi^\munu a_\nu,\quad
        t^{\{\munu\}}_{\perp} \equiv \Xi^\munu_{\alpha\beta}t^{\alpha\beta},
  \nonumber\\
  && \dot{a}^{\{\mu\}} \equiv \Xi^\munu \dot{a}_\nu,\quad
  \dot{t}^{\{\munu\}}_{\perp} \equiv \Xi^\munu_{\alpha\beta}\dot{t}^{\alpha\beta}
\end{eqnarray}
for the spatially transverse (in the LRF) components of a vector $a^\mu$ or its LRF time derivative $\dot{a}^\mu$ and the spatially transverse and traceless part of a tensor $t^\munu$ or its LRF time derivative $\dot{t}^\munu$, as well as
\be
\label{eq26}
  \int_p = \frac{g}{(2\pi)^3} \int d^4p \, 2\Theta(p^0) \delta(p^2-m^2) = \frac{g}{(2\pi)^3} \int \frac{d^3p}{E_p}
\ee
for the Lorentz-invariant momentum space integral, with $g$ being a degeneracy factor counting the number of quantum states allowed for a particle with on-shell momentum $p^\mu$, and $\Theta(p^0)$ denoting the Heaviside step function.  

After moving the time derivative $D$ on the r.h.s. under the integral until it hits the distribution function $f_a$ or $\dft$, we use the decomposition $f=f_a+\dft$ together with 
\be
\label{eq27}
  \del_\mu = u_\mu D + z_\mu D_z + \nabla_{\perp\mu}
\ee
to rewrite the Boltzmann-Vlasov equation \eqref{eq13} in the form
\be
\label{eq28}
\begin{split}
  \dot{f}_a + \delta\dot{\tilde f} &= \frac{C[f]- m \, \del^\mu m \, \del_\mu^{(p)} f}{\up}  \\
                                                &+  \frac{\mzp D_z f_a - \pperp \nabla_{\perp\mu} f_a}{\up} \\
                                                &+ \frac{\mzp D_z \dft - \pperp \nabla_{\perp\mu} \dft}{\up} \,.
\end{split}
\ee
Closing this equation requires an approximation for $\dft$. We here use the 14-moment approximation. 
 
%%%%%%%%%%%%%%%%%%%%%%%%%%%%%%%%%%%%%%%%%%%%%%%%%%%%%%%%%%
\subsection{The 14-moment approximation}
\label{sec3c}
%%%%%%%%%%%%%%%%%%%%%%%%%%%%%%%%%%%%%%%%%%%%%%%%%%%%%%%%%%

The 14-moment approximation derives its name from approximating $\dft$ in terms of its 14 momentum moments with $p^\mu$ and $p^\mu p^\nu$ (where the moment with the linear combination $p^\mu p^\mu g_\munu=m^2$ as weight function is equivalent with the moment taken with weight 1) \cite{CPA:CPA3160020403, Israel:1979wp}. In our case the choice of the Landau frame, together with the generalized matching conditions (\ref{eq19}), (\ref{eq20}), and (\ref{eq22}a,b) and the absence of diffusion currents related to conserved charges, eliminate ten of these moments, leaving only four independent moments to construct $\dft$. These need to be matched to the residual dissipative flows $\Wperp$ and $\piperp$, which each have two degrees of freedom. The 14-moment approximation for $\dft$ can thus be written as \cite{Molnar:2016vvu}
\be
\label{eq29}
   \frac{\dft}{f_a \bar{f}_a}(x,p) = c_{\perp}^{\{\mu\}}(x)\bigl(-z(x){\cdot}p\bigr) p_{\{\mu\}} 
                                                              + c_{\perp}^{\{\munu\}}(x) p_{\{\mu} \, p_{\nu\}},
\ee
where $\bar{f}_a{\,=\,}1{-}\Theta f_a$. The coefficients $c_{\perp}^{\{\mu\}}(x)$ and $c_{\perp}^{\{\munu\}}(x)$ are computed by substituting Eq.~\eqref{eq29} into the kinetic theory definitions of $\Wperp$ and $\piperp$,
\bs
\label{eq30}
\beal
&\Wperp = \int_p  \mzp p^{\{\mu\}} \dft \,, \\
&\piperp = \int_p p^{\{\mu}p^{\nu\}} \dft \,,
\end{align}
\es
and decoupling the resulting set of linear equations:
\be   
\label{eq31}
\begin{aligned}
 c_{\perp}^{\{\mu\}} = - \frac{\Wperp}{\J_{4210}}, \idd & \idd c_{\perp}^{\{\munu\}} = \frac{\piperp}{2\J_{4020}}. 
\end{aligned}
\ee
The anisotropic integrals $\J_{nrqs}$ appearing in these expressions are defined in Eq.~(\ref{eqA2}). As expected, the coefficients are directly proportional to the residual dissipative flows: 
\be
\label{eq32}
  \dft = \left( -\mzp\, \frac{p_{\{\mu\}}\Wperp}{\J_{4210}} 
        + \frac{p_{\{\mu}\,p_{\nu\}}\piperp}{2 \J_{4020}}\right) f_a \bar{f}_a \,.
\ee
%

%%%%%%%%%%%%%%%%%%%%%%%%%%%%%%%%%%%%%%%%%%%%%%%%%%%%%%%%%%
\subsection{Relaxation equations II}
\label{sec3d}
%%%%%%%%%%%%%%%%%%%%%%%%%%%%%%%%%%%%%%%%%%%%%%%%%%%%%%%%%%

Substituting the 14-moment approximation (\ref{eq32}) for $\dft$ into Eqs.~(\ref{eq24}) and (\ref{eq28}), simplifying some of the~resulting \\ \vfill \noindent terms by integrating by parts, and enforcing the generalized matching conditions, some algebra yields the following dissipative relaxation equations:\footnote{%
	In deriving these equations one encounters terms involving the comoving time derivative 
	$\dot{m}$ of the temperature-dependent quasiparticle mass that arise from the second term 
	on the l.h.s. of the Boltzmann-Vlasov equation (\ref{eq13}). We eliminate them by using the 
	chain rule $\dot{m} = (dm/dT)\,(dT/d\ene)\,\dot{\ene}$ where we take $dm/dT$ as external input 
	from the quasiparticle model discussed below, evaluate the derivative $dT/d\ene= (d\ene/dT)^{-1} 
	= c_s^2 T / (\ene+\Peq)$ from the lattice QCD EOS, and use Eq.\,(\ref{eq10}) for $\dot{\ene}$.
	Equations~(\ref{eq33})--(\ref{eq36}) (together with the transport coefficients listed in 
	Appendix~\ref{appa}) are found after combining the terms on the r.h.s. of Eq.~(\ref{eq10}) with
	other terms involving the same dissipative forces. In doing so we neglect contributions of 
	$\mathcal{O}(\text{Kn}\,\tilde{\text{R}}^{-1}_i \tilde{\text{R}}^{-1}_j)$, where $\tilde{\text{R}}^{-1}_i$ 
	is the residual inverse Reynolds number associated with the residual components 
	$\Wperp$ and $\piperp$.
	\label{fn5}
	}
\begin{widetext} 
\begin{eqnarray} 
\label{eq33}
   \dot{\mathcal{P}}^{(k)}_L &=&  
   - \frac{\bar{\mathcal{P}}^{(k)}{-}\Peq^{(k)}}{\tau_\Pi}
   - \frac{\PL^{(k)}{-}\Pperp^{(k)}}{3\tau_\pi / 2} + \bar{\zeta}^{L(k)}_z z_\mu D_z u^\mu 
   + \bar{\zeta}^{L(k)}_\perp \theta_\perp - 2\Wperp \dot{z}_\mu 
   + \bar{\lambda}^{L(k)}_{Wu} \Wperp D_z u_\mu 
   \nonumber\\
   &&+ \bar{\lambda}^{L(k)}_{W\perp} \Wperp z_\nu \nabla_{\perp\mu} u^\nu 
   - \bar{\lambda}^{L(k)}_{\pi} \piperp \sigma_{\perp,\munu} ,
\\
\label{eq34}
  \dot{\mathcal{P}}^{(k)}_\perp &=&  
  - \frac{\bar{\mathcal{P}}^{(k)}{-}\Peq^{(k)}}{\tau_\Pi} + \frac{\PL^{(k)}{-}\Pperp^{(k)}}{3\tau_\pi} 
  + \bar{\zeta}^{\perp(k)}_z z_\mu D_z u^\mu + \bar{\zeta}^{\perp(k)}_\perp \theta_\perp 
  + \Wperp \dot{z}_\mu + \bar{\lambda}^{\perp(k)}_{Wu} \Wperp D_z u_\mu
  \nonumber\\ 
  &&- \bar{\lambda}^{\perp(k)}_{W\perp} \Wperp z_\nu \nabla_{\perp\mu} u^\nu 
  + \bar{\lambda}^{\perp(k)}_{\pi} \piperp \sigma_{\perp,\munu} ,
\\
\label{eq35}
   \dot{W}^{\{\mu\}}_{\perp z} &=&  
   - \frac{\Wperp}{\tau_\pi} + 2\bar{\eta}^W_u \Xi^\munu D_z u_\nu 
   - 2\bar{\eta}^W_\perp z_\nu \nabla_\perp^\mu u^\nu 
   - \big(\bar{\tau}^W_z \Xi^\munu + \piperp\big) \dot{z}_\nu 
   + \bar{\delta}^W_W \Wperp \theta_\perp 
   \nonumber\\
   && - \bar{\lambda}^W_{W u} \Wperp  z_\nu D_z u^\nu  
   + \bar{\lambda}^W_{W \perp} \sigma_\perp^\munu  W_{\perp z, \nu}
   + \omega_\perp^\munu W_{\perp z, \nu} +  \bar{\lambda}^W_{\pi u} \piperp D_z u_\nu 
   -  \bar{\lambda}^W_{\pi \perp} \piperp z_\alpha \nabperp_\nu u^\alpha,
\\
\label{eq36}
   \dot{\pi}^{\{\munu\}}_{\perp} &=&  
   - \frac{\piperp}{\tau_\pi} + 2 \bar{\eta}_\perp \sigma_\perp^\munu 
   - 2 W_{\perp z}^{\{\mu} \dot{z}^{\nu\}} - \bar{\delta}^\pi_\pi \, \piperp \theta_\perp 
   - \bar{\tau}^\pi_\pi  \, \pi_\perp^{\alpha \{\mu} \sigma^{\nu\}}_{\perp,\alpha} 
   + 2 \pi_\perp^{\alpha \{\mu} \omega^{\nu\}}_{\perp,\alpha} + \bar{\lambda}^\pi_\pi \, \piperp z_\alpha D_z u^\alpha  
   \nonumber\\
   && - \bar{\lambda}^\pi_{W u} \, W_{\perp z}^{\{\mu} D_z u^{\nu\}} 
   + \bar{\lambda}^\pi_{W \perp} \, W_{\perp z}^{\{\mu} z_\alpha \nabla_\perp^{\nu\}} u^\alpha.
\end{eqnarray}
\end{widetext}
Here $\Pavg^{(k)} = \frac{1}{3}(\PL^{(k)}{+}2\Pperp^{(k)})$ is the average pressure as given by kinetic theory, and $\omega_\perp^\munu{\,\equiv\,}\Xi^\mu_\alpha \Xi^\nu_\beta\, \del^{[\beta} u^{\alpha]}$ is the transverse vorticity tensor. 

The structure of Eqs.\,(\ref{eq33})--(\ref{eq36}) is simpler than that of the corresponding equations derived in \cite{Molnar:2016vvu}, not only by the absence of terms coupling to the conserved charge and diffusion currents (which only reflects the simplifying assumptions made here), but also as a result of imposing the generalized Landau matching conditions (\ref{eq19}), (\ref{eq20}), and (\ref{eq22}a,b) which optimizes the evolution of the anisotropy parameters in $f_a$ and thus removes additional terms needed in \cite{Molnar:2016vvu} to correct their evolution if not chosen optimally in the first place. 

The transport coefficients appearing on the right hand sides of Eqs.~(\ref{eq33})-(\ref{eq36}) are labeled following as much as possible the convention established in Ref.~\cite{Molnar:2016vvu}. Except for the relaxation times they are given in Appendix~\ref{appb}.\footnote{%
    Please note the superscripts $(k)$ on the transport coefficients appearing on the right hand sides of 
    Eqs.~(\ref{eq33}) and (\ref{eq34}). They reflect the fact that these control the evolution of the kinetic
    part of the longitudinal and transverse pressures. For the quasiparticle model introduced in the next Section
    an additional mean field enters which modifies these pressures and transport coefficients. The modified
    expressions will be denoted without the superscript.
    }
Generically they involve the ``anisotropic thermodynamic integrals'' over the anisotropic distribution function $f_a$ given in Appendix~\ref{appa}. Their validity, as well as the validity of the specific relations between some of these transport coefficients listed in Appendix~\ref{appb}, depends on the applicability of relativistic kinetic theory of a gas of weakly-interacting quasiparticles as the underlying microscopic theory, which is not guaranteed for quark-gluon plasma. Their generalization to a realistic microscopic theory of QCD medium dynamics requires much additional work. We will here use the expressions given in the Appendices as order-of-magnitude estimates and placeholders for future more realistic sets of transport coefficients. 

Equations~(\ref{eq33})-(\ref{eq36}) also involve two relaxation times, $\tau_\pi$ and $\tau_\Pi$. $\tau_\Pi$ controls the relaxation of the kinetic bulk viscous pressure $\Pavg^{(k)}{-}\Peq^{(k)}$ (see Eq.~(\ref{eq8})) whereas the shear relaxation time $\tau_\pi$ drives the relaxation of both the large shear stress component $\PL^{(k)}{-}\Pperp^{(k)}$ and the smaller ones described by $\Wperp$ and $\piperp$. That all shear stress components have the same relaxation time even if some of them become large is a model assumption that may be corrected in future improved calculations of the transport coefficients for strongly anisotropically expanding QGP. 

Formally, the relaxation times arise from a linearization of the collision term around the local equilibrium distribution $f_\eq$ (with temperature computed from the energy density):
\be
\label{eq37}
  C[f]= -\frac{p\cdot u(x)}{\tau_r(x)}\Bigl(f(x,p)-f_\eq\Bigl(\frac{p{\cdot}u(x)}{T(x)}\Bigr)\Bigr).
\ee
Literal use of this Relaxation Time Approximation (RTA) \cite{Anderson_Witting_1974} gives $\tau_\pi = \tau_\Pi = \tau_r$. However, strong coupling in the quark-gluon plasma in the temperature regime just above  the quark-hadron phase transition, as well as critical behavior near that phase transition, lead to very different temperature dependences of the bulk and shear viscosities and their associated relaxation times in QCD, especially around $T_c$ \cite{Csernai:2006zz, Paech:2006st, Arnold:2006fz, Kharzeev:2007wb, Karsch:2007jc}. In particular, the bulk relaxation time $\tau_\Pi$ is expected to be affected by ``critical slowing down'' \cite{Arnold:2006fz, Berdnikov:1999ph, Song:2009rh}, i.e. it should exhibit a strong peak near $T_c$. Since large bulk viscous effects near $T_c$ are one of the main motivations for our work here, we feel compelled to account for them by introducing two different relaxation times $\tau_\pi$ and $\tau_\Pi$, and tying them to phenomenologically parametrized shear and bulk viscosities $\eta$ and $\zeta$ by postulating the standard kinetic theory relations \cite{Denicol:2012cn}
\be
\label{eq38}
   \tau_\pi = \eta / \beta_\pi \, \quad
   \tau_\Pi = \zeta / \beta_\Pi.
\ee
The (temperature dependent) isotropic thermodynamic integrals $\beta_\pi$ and $\beta_\Pi$ appearing in these relations are given further below in Eq.~(\ref{eq:beta}). The viscosities $\eta$ and $\zeta$ are transport parameters that occur in standard ``isotropic'' dissipative fluid dynamics -- they appear here through the relaxation times $\tau_\pi$ and $\tau_\Pi$. When comparing anisotropic with standard dissipative fluid dynamics further below in Sec.~\ref{sec5} we will do so by using the same functions $\tau_\pi$ and $\tau_\Pi$ in both approaches. 

%%%%%%%%%%%%%%%%%%%%%%%%%%%%%%%%%%%%%%%%%%%%%%%%%%%%%%%%%
\section{Anisotropic equation of state}
\label{sec4}
%%%%%%%%%%%%%%%%%%%%%%%%%%%%%%%%%%%%%%%%%%%%%%%%%%%%%%%%%

While the relaxation equations~\eqref{eq33}--\eqref{eq36} were derived from the Boltzmann equation, the equations remain structurally unchanged for strongly coupled fluids. They are purely macroscopic, i.e. all terms on the r.h.s.  have the form of some macroscopic driving force (proportional to the Knudsen or inverse Reynolds numbers or products thereof) multiplied by some transport coefficient. The kinetic origin of these equations is hidden in these transport coefficients. Applying the equations to strongly coupled fluids requires only that these transport coefficients, along with the equation of state relating the energy density and equilibrium pressure, are swapped out accordingly.

For the time being most of the transport coefficients of hot and dense QCD matter are still essentially unknown. While the shear and bulk viscosities will be taken as parameters whose functional forms are modeled phenomenologically and whose overall magnitudes are to be fitted to experimental observables, the remaining transport coefficients will be approximated using kinetic theory, for reasons of consistency with our derivation of the evolution equations. Their evaluation requires microscopic kinetic inputs, namely the parameters $(\Lambda, \alpha_\perp, \alpha_L)$ characterizing the anisotropic distribution function $f_a$, the particle mass $m$, and also the temperature $T$. However, for the QGP equation of state, which is very precisely known from lattice QCD calculations \cite{Borsanyi:2010cj, Bazavov:2014pvz}, we want to use first-principles theoretical input. 

In this Section we discuss how to consistently incorporate such direct information from QCD into a hydrodynamic framework that was originally derived from a kinetic theory with a very different EOS. We will call this procedure ``integrating the lattice QCD EOS with some kinetic framework''. We introduce a parametric model for an anisotropic equation of state that allows the anisotropic hydrodynamic equations, including the dissipative relaxation equations for the longitudinal and transverse pressures and the remaining shear stress components, to be solved on a purely macroscopic level. This differs from earlier implementations of the framework which relied on the solution of dynamical evolution equations of the microscopic kinetic parameters $(\Lambda, \alpha_\perp, \alpha_L, m)$ \cite{Martinez:2010sc,Martinez:2012tu,Molnar:2016gwq, Alqahtani:2016rth, Alqahtani:2017jwl, Alqahtani:2017tnq} (which, for the case of QCD, are not really well-defined). However, since we will need these microscopic parameters for the calculation of those transport coefficients that we compute from kinetic theory (a temporary necessity that will disappear as soon as ways have been found to calculate these transport coefficients directly from QCD), we determine them from the macroscopic hydrodynamic quantities at the end of each time evolution step, using our parametric model for the anisotropic EOS.\footnote{%
	Note that the parametric model is not used for the equilibrium EOS $\Peq(\ene)$ (for which we 
	take state-of-the-art lattice QCD results) but only to parametrize the dissipative deviations of the
	longitudinal and transverse pressures from $\Peq(\ene)$, as well as for the calculation of the remaining 
	transport coefficients.
	}   

To construct this parametric model we follow Refs.~\cite{Alqahtani:2015qja, Alqahtani:2016rth, Tinti:2016bav} and parametrize the response of the pressure anisotropy and the bulk viscous pressure to anisotropic expansion using a quasiparticle EOS. The quasiparticles have a temperature-dependent mass that is chosen such that a weakly-interacting gas of these particles accurately mimics the QCD EOS. The transport coefficients are then worked out in this kinetic theory.\footnote{%
	Note that an accurate description of the equation of state does not imply by any means that the 
	kinetic theory also predicts the correct transport properties of the medium. In all likelihood it doesn't.
	}
It is well known \cite{Biro:1990vj, Gorenstein:1995vm, Peshier:1995ty} that for thermodynamic consistency such an approach requires the introduction of a mean field $B$ whose temperature dependence in equilibrium generates the temperature dependence of the quasiparticle's effective mass. It also receives additional dissipative corrections out of equilibrium \cite{Tinti:2016bav}.

%%%%%%%%%%%%%%%%%%%%%%%%%%%%%%%%%%%%%%%%%%%%%%%%%%%%%%%
\subsection{Integrating the lattice QCD EOS with a quasiparticle EOS}
\label{sec4a}
%%%%%%%%%%%%%%%%%%%%%%%%%%%%%%%%%%%%%%%%%%%%%%%%%%%%%%%

The key question that needs to be addressed in anisotropic hydrodynamics is how much pressure anisotropy and bulk viscous pressure is generated by a given hydrodynamic expansion rate and its anisotropy. These are the two  largest and most important dissipative effects in our approach. The answer to this question depends on the microscopic properties of the medium. For QCD matter this response is presently not known. It is, however, a key ingredient in the hydrodynamic evolution model. In this subsection we model this response by that of a weakly interacting gas of quasiparticles with a medium-dependent mass $m(T)$. Within this model we can associate (within certain limits) any given deviations of the longitudinal and transverse pressures $\PL$ and $\Pperp$ from the equilibrium pressure $\Peq(\ene)$ with specific values for the microscopic parameters $\bigl(\Lambda, \bm{\alpha}, m\bigl(T)\bigr)\bigr)$ describing the anisotropic quasiparticle distribution function $f_a$. These values can then be used to compute the kinetic theory values for the transport coefficients. So while the equilibrium pressure is described by the full QCD EOS from lattice QCD, the dissipative deviations of $\PL$ and $\Pperp$ from the equilibrium pressure are interpreted microscopically within a weakly interacting gas of massive Boltzmann particles. As we solve the hydrodynamic equations (\ref{eq10})--(\ref{eq12}) together with the dissipative relaxation equations (\ref{eq33})--(\ref{eq36}), we interpret the resulting deviations from local equilibrium within the quasiparticle model by writing
\bs
\label{eq39}
\beal
  0 &= \ene^{(q)} - \ene_{\eq}^{(q)}\bigl(\ene\bigr), \\
  \PL - \Peq(\ene) &= \PL^{(q)} - \mathcal{P}_{\eq}^{(q)}\bigl(\ene\bigr), \\
  \Pperp - \Peq(\ene) &= \Pperp^{(q)} - \mathcal{P}_{\eq}^{(q)}\bigl(\ene\bigr).
\end{align}
\es
Here the superscript $(q)$ stands for ``quasiparticle model''. The zero on the l.h.s. of the first of these equations reflects the Landau matching condition $\ene=\ene(T)$ to the lattice QCD energy density, which also provides us with the temperature $T$ at which the quasiparticle mass $m(T)$ and equilibrium mean field $B_\eq(T)$ (see below) are evaluated. In the quasiparticle model the hydrodynamic quantities on the r.h.s. of~\eqref{eq39} consist of kinetic and mean field contributions  \cite{Alqahtani:2015qja}
\bs
\label{eq40}
\beal
  \ene^{(q)} &= \ene^{(k)} + B,\\
  \PL^{(q)} &= \PL^{(k)} - B, \\
   \Pperp^{(q)} &= \Pperp^{(k)} - B. 
\end{align}
\es
The kinetic contributions are obtained from Eqs.~(\ref{eq23}b,c,d):
\bs
\label{eq41}
\beal
&\ene^{(k)}\bigl(\Lambda,\bm{\alpha}; m(T)\bigr)  = \I_{2000}\bigl(\Lambda,\bm{\alpha}; m(T)\bigr), \\
&\PL^{(k)}\bigl(\Lambda,\bm{\alpha}; m(T)\bigr)  = \I_{2200}\bigl(\Lambda,\bm{\alpha}; m(T)\bigr), \\
&\Pperp^{(k)}\bigl(\Lambda,\bm{\alpha}; m(T)\bigr)  = \I_{2010}\bigl(\Lambda,\bm{\alpha}; m(T)\bigr),
\end{align}
\es
where $T = T(\ene)$. The mean field $B$ consists of an equilibrium part $B_\eq$ and a dissipative correction $\delta B$:
\be
\label{eq42}
 B = B_\eq(T) + \delta B.
\ee
By Landau matching, the total quasiparticle energy density $\ene^{(q)}$ is fixed to its equilibrium value:
\be
\label{eq43}
  \ene_\eq^{(q)}(\ene) = \ene^{(k)}_\eq(T) + B_{\eq}(T),
\ee
where $\ene^{(k)}_\eq(T) = \I_{2000}\bigl(T,\bm{1};m(T)\bigr)$. The Landau matching condition (\ref{eq39}a) can then be rewritten as
\be
\label{eq44}
  \I_{2000}\bigl(\Lambda,\bm{\alpha}; m(T)\bigr) = \I_{2000}\bigl(T,\bm{1};m(T)\bigr) - \delta B.
\ee
This establishes a relation between the temperature and the kinetic theory parameters, provided that $\delta B$ is determined. In the equilibrium limit, the quasiparticle pressure is
\be
\label{eq45}
   \Peq^{(q)}(\ene) = \Peq^{(k)}(T)  - B_\eq(T)
\ee
where $\Peq^{(k)}(T) = \I_{2200}\bigl(T,\bm{1};m(T)\bigr)$. The equilibrium terms in (\ref{eq43}) and (\ref{eq45}) are all functions of temperature. 

For simplicity we assume that the quasiparticles have Boltzmann statistics ($\Theta{\,=\,}0$). To ensure that at asymptotically high temperature the equilibrium pressure and energy density of this Boltzmann gas approach the corresponding values of a quark-gluon gas with $2(N_c^2{-}1)$ bosonic and $4N_cN_f$ fermionic degrees of freedom, we normalize them by applying to the quasiparticle distribution function a degeneracy factor
\be
\label{eq46}
  g = \Bigl(2(N_c^2{-}1) + 4 N_c N_f {\textstyle\frac{7}{8}}\Bigr)\, \frac{\pi^4}{90}
\ee
with $N_c = 3$ colors and  $N_f = 3$ flavors, counting $u$, $d$, and $s$ quarks only (heavier flavors are exponentially suppressed in the phenomenologically interesting temperature range and are therefore neglected). This degeneracy factor is part of the momentum integration measure $\int_p$ in the definition (\ref{eq26}).

%%%%%%%%%%%%%%%%%%%% Fig. 1 %%%%%%%%%%%%%%%%%%%%%%%%%%%%%%%%%
\begin{figure}[t]
\centering
\includegraphics[]{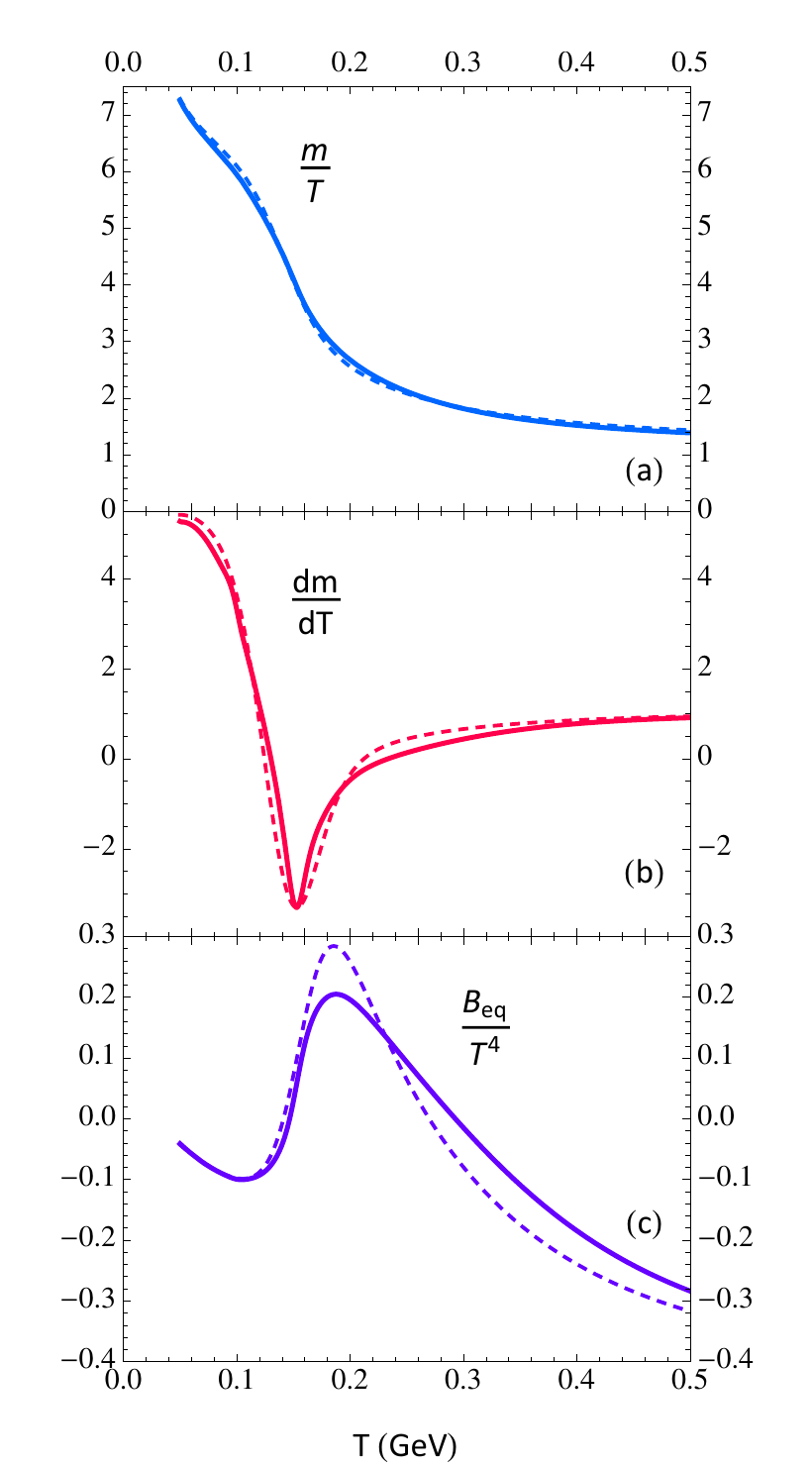}
\caption{(Color online) 
	The normalized quasiparticle mass $z=m/T$ (a), the derivative $dm/dT$ (b) and the equilibrium 
	mean field $B_\eq/T^4$ (c) obtained from Eqs.~(\ref{eq47}) and (\ref{eq48}) using a state-of-the-art
	lattice QCD EOS compiled by the BEST Collaboration \cite{Parotto:2018pwk} (solid lines) and 	
	from earlier lattice QCD results obtained by the Wuppertal-Budapest Collaboration 
	\cite{Borsanyi:2010cj, Alqahtani:2015qja} (dashed lines), shown as functions of temperature $T$. 
	In practice, the equilibrium mean field in panel (c) is obtained as the difference between the 
	equilibrium kinetic and lattice QCD pressures, $B_\eq=\Peq^{(k)}{-}\Peq^{(q)}=\Peq^{(k)}{-}\Peq$ 
	(see Eq.~(\ref{eq45})).
        \label{F1}}
\end{figure}

%%%%%%%%%%%%%%%%%%%%%%%%%%%%%%%%%%%%%%%%%%%%%%%%%%%%%%%%

The thermal quasiparticle mass $m(T)$ is chosen such that the equilibrium pressure $\Peq^{(q)}(\ene)$ and energy density $\ene_\eq^{(q)}(\ene)$ of the quasiparticle model agree with their lattice QCD counterparts. Technically this is done by expressing the lattice QCD entropy density $\mathcal{S}$ in terms of the corresponding kinetic theory expression $\mathcal{S}^{(q)}$ for a gas of quasiparticles with mass $m(T)$ and Boltzmann statistics \cite{Alqahtani:2015qja}: 
\begin{eqnarray}
\label{eq47}
   \mathcal{S} &=& \frac{\ene+\Peq}{T}
\\\nonumber
   &=& \frac{\ene_\eq^{(q)}+\Peq^{(q)}}{T}  = \frac{\ene_\eq^{(k)}+\Peq^{(k)}}{T}  =  \frac{g T^3 z}{2 \pi^2}  K_3(z),
\end{eqnarray}
where $K_n(z)$ is the modified Bessel function with $z = m(T)/T$. For thermodynamic consistency the right hand side must satisfy $\mathcal{S}^{(q)} = d\Peq^{(q)}/dT$, which is ensured by setting 
\be
\label{eq48}
   B_\eq(T) = - \frac{g}{2\pi^2}\int^T_0 d\hat{T}\, {\hat{T}}^3 {\hat{z}}^2 K_1(\hat{z})\, \frac{dm(\hat{T})}{d\hat{T}}
\ee
where $\hat{z}= m(\hat{T})/\hat{T}$. Plots of the quasiparticle mass-to-temperature ratio $z$ and equilibrium mean field $B_\eq$ as functions of $T$, using a 2010 lattice QCD EOS obtained by the Wuppertal-Budapest Collaboration \cite{Borsanyi:2010cj}, can be found in \cite{Alqahtani:2015qja}. Here we use the state-of-the-art QCD EOS compiled by the Beam Energy Scan Theory (BEST) Collaboration \cite{Parotto:2018pwk}. The resulting slightly modified temperature dependences of $z(T)$, $dm(T)/dT$ and $B_\eq(T)$ are shown in Fig.~\ref{F1} as solid lines (together with the earlier results from Ref.~\cite{Alqahtani:2015qja} shown as dashed lines).

Equation~(\ref{eq48}) determines the mean field in equilibrium. Out of equilibrium it receives a non-equilibrium correction $\delta B$ \cite{Tinti:2016bav}.  As shown in \cite{Tinti:2016bav}, thermodynamic consistency and energy-momentum conservation can be used to derive from the Boltzmann equation the following general evolution equation for $B$: 
\be
\label{eq49}
  \dot{B} + m \dot m \int_p f + \int_p (\up)\, C[f] = 0. 
\ee
By Landau matching the non-equilibrium correction to the quasiparticle energy density $\ene^{(q)} = \ene^{(k)}{+}B$ must vanish, hence
\be
\label{eq50}
  B = B_\eq - \int_p (\up)^2 \delta f 
\ee
where $\delta f=f{-}f_\eq$. Substituting $f = f_a{+}\dft$ and using the relaxation time approximation
\be
\label{eq51}
   C[f] \approx - \frac{(\up)\,\delta f}{\tau_r}
\ee
equation~\eqref{eq49} takes the form
\be
\label{eq52}
  \dot{B} = -\frac{B{-}B_\eq}{\tau_\Pi} - \frac{\dot{m}}{m} \Big(\ene^{(k)}{-}2\Pperp^{(k)}{-}\PL^{(k)} \Big).
\ee
Note that the expression in the parentheses is the trace of the kinetic contribution to the energy-momentum tensor $T^\munu$. Since the non-equilibrium component of the mean field $\delta B = B - B_\eq$ contributes to the bulk viscous pressure, we have replaced in Eq.~(\ref{eq52}) the relaxation time $\tau_r$ by the bulk relaxation time $\tau_\Pi$. The time derivative of the thermal mass can be expressed in terms of the energy conservation law~\eqref{eq10} using the chain rule. 

Although Eq.~(\ref{eq50}) shows that $B$ is not an independent quantity, we find it most straightforward to use  Eq.~\eqref{eq52} to propagate the mean field $B$ dynamically. It does not directly enter the evolution equations for the components of the energy momentum tensor as an independent variable, but is only needed for the model interpretation of the pressure anisotropy and bulk viscous pressure (which are hydrodynamic outputs) in terms of the microscopic parameters $(\Lambda,\bm{\alpha})$ needed for computing the transport coefficients in Appendix~\ref{appb}. We use Eqs.~(\ref{eq10}), (\ref{eq33}), (\ref{eq34}) and (\ref{eq52}) to evolve $\ene$, $\PL^{(k)}$, $\Pperp^{(k)}$ and $B$. The physical pressures are obtained from $\PL=\PL^{(k)}{-}B$ and $\Pperp=\Pperp^{(k)}{-}B$. From $\ene$ we determine $T$ using the lattice EOS, and thus we know $m(T)$. Then we rewrite our anisotropic equation of state model (\ref{eq39}) as%
\bs
\label{eq53}
\beal
  \ene - B &= 
  \I_{2000}\bigl(\Lambda, \bm{\alpha}, m(T)\bigr) , \\
  \PL + B &= 
  \I_{2200}\bigl(\Lambda, \bm{\alpha}, m(T)\bigr), \\
  \Pperp + B &= 
  \I_{2010}\bigl(\Lambda, \bm{\alpha}, m(T)\bigr),
\end{align}
\es
solve these equations for the anisotropy parameters $(\Lambda, \alpha_\perp,\alpha_L)$, and compute the transport coefficients. 

Of course, the values $\bigl(\Lambda, \bm{\alpha}, m\bigr)$ associated in this way with $\PL$, $\Pperp$ and $\ene$ at any point of the hydrodynamic space-time grid are model dependent, and a different parametrization of the lattice QCD EOS in terms of quasiparticles (for example, as a mixture of different types of quasiparticles with different quantum statistical properties, degeneracy factors and masses) would yield different results. For example, we have tried (and abandoned) an alternate approach where we used a weakly interacting Boltzmann gas of particles with a fixed mass to interpret the pressure anisotropy and bulk viscous pressure in terms of microscopic parameters $(\Lambda,\bm{\alpha};m)$. In that case we were unable to find solutions at early times where the strong longitudinal expansion leads to negative bulk viscous pressures large enough that no valid choice of microscopic parameters can reproduce this in the kinetic model theory. In the quasiparticle model we can partially absorb this with an out-of-equilibrium mean field contribution.

With the anisotropic EOS model (\ref{eq53}) we can finally write down the equations of motion for the total pressures $\PL$ and $\Pperp$, by combining Eqs.~(\ref{eq33},\ref{eq34}) with Eq.~(\ref{eq52}):
\begin{widetext}
\begin{eqnarray} 
\label{eq54}
   \dot{\mathcal{P}}_L &=& - \frac{\bar{\mathcal{P}}{-}\Peq}{\tau_\Pi}
   - \frac{\PL{-}\Pperp}{3\tau_\pi / 2} + \bar{\zeta}^L_z z_\mu D_z u^\mu + \bar{\zeta}^L_\perp \theta_\perp
   - 2\Wperp \dot{z}_\mu + \bar{\lambda}^L_{Wu} \Wperp D_z u_\mu
   + \bar{\lambda}^L_{W\perp} \Wperp z_\nu \nabla_{\perp\mu} u^\nu  
   - \bar{\lambda}^L_{\pi} \piperp \sigma_{\perp,\munu} ,\qquad
\\
\label{eq55}
  \dot{\mathcal{P}}_\perp &=& - \frac{\bar{\mathcal{P}}{-}\Peq}{\tau_\Pi} 
  + \frac{\PL{-}\Pperp}{3\tau_\pi} + \bar{\zeta}^\perp_z z_\mu D_z u^\mu + \bar{\zeta}^\perp_\perp \theta_\perp 
  + \Wperp \dot{z}_\mu + \bar{\lambda}^\perp_{Wu} \Wperp D_z u_\mu 
   - \bar{\lambda}^\perp_{W\perp} \Wperp z_\nu \nabla_{\perp\mu} u^\nu 
  + \bar{\lambda}^\perp_{\pi} \piperp \sigma_{\perp,\munu}.
\end{eqnarray}
\end{widetext}
Here we redefined the transport coefficients for the longitudinal and transverse pressures as detailed in Appendix~\ref{appc}. This completes our formalism for nonconformal anisotropic hydrodynamics, where the equations of motion (\ref{eq10})-(\ref{eq12}), (\ref{eq35}), (\ref{eq36}), (\ref{eq54}) and (\ref{eq55}) are purely macroscopic and structurally independent of the underlying microscopic physics, while the transport coefficients are evaluated with our specific quasiparticle kinetic model for the anisotropic equation of state.

%%%%%%%%%%%%%%%%%%%%%%%%%%%%%%%%%%%%%%%%%%%%%%%%%%%%%%%%%%
\subsection{Reconstructing energy density and fluid velocity}
\label{sec4b}
%%%%%%%%%%%%%%%%%%%%%%%%%%%%%%%%%%%%%%%%%%%%%%%%%%%%%%%%%%

Most numerical codes developed for heavy-ion collisions (including ours) solve the energy-momentum conservation laws (\ref{eq9}) on a fixed ``Eulerian'' space-time grid instead of the LRF projected conservation equations (\ref{eq10})--(\ref{eq12}). The reason for this is the existence of powerful flux-corrected evolution algorithms for conservation laws of the type (\ref{eq9}) \cite{Kurganov:2000:NHC:346416.346437, Schenke:2010nt, Bazow:2016yra}. To be able to use these algorithms one writes Eqs.~(\ref{eq9}) in conserved flux form, which for Milne coordinates $x^\mu = (\tau,x,y,\eta)$ reads
\be
\label{eq55a}
\partial_\tau T^{\tau\mu} + \partial_k(v^k T^{\tau \mu}) = J^\mu.
\ee
The source term $J^\mu$ on the r.h.s. includes both geometric (Christoffel) terms arising from the curvilinear nature of the coordinate system and the dissipative fluxes \cite{Bazow:2016yra}. 

Running the evolution algorithm for one time step thus produces updated values for the first row of the energy-momentum tensor and the dissipative flows encoded in $\PL$, $\Pperp$, $\Wperp$ and $\piperp$, as well as the mean field $B$. For the evaluation of the EOS and the calculation of the transport coefficients as described in the preceding subsection we need, however, the LRF energy density $\ene$, and to compute the source term $J^\mu$ we need the fluid four-velocity $u^\mu$. How to obtain these from the updated output of the evolution code at the end of a time step is described in this subsection.  

We start by writing the first row of $T^\munu$ as
\be
\label{eq55b}
   T^{\tau\mu} = (\ene{+}\Pperp)u^\tau\um - \Pperp g^{\tau\mu} + \Delta\mathcal{P} z^\tau z^\mu  
                       + 2 W^{(\tau}_{\perp z} z^{\mu)} + \pi_\perp^{\tau\mu}
\ee
where $g^\munu = \diag(1,-1,-1,-1/\tau^2)$ is the metric tensor and $\Delta\mathcal{P} \equiv \PL{-}\Pperp$. The space-like basis vector $z^\mu$ can be parameterized as \cite{Molnar:2016vvu}
\be
\label{eq56}
  z^\mu = \frac{1}{\sqrt{1{+}u_\perp^2}}(\tau u^\eta, 0, 0, u^\tau/\tau) 
\ee
where $u_\perp = \sqrt{u_x^2{+}u_y^2}$ is the magnitude of the transverse four-velocity. 

Next, we observe that the term $2W^{(\tau}_{\perp z} z^{\mu)}$ depends (through the components of $z^\mu$) on $u^\tau$ and $u^\eta$. It cannot be subtracted from $T^{\tau\mu}$ until a relation between $u^\tau$ and $u^\eta$ is found. To this end let us construct from known quantities the vector $K^\mu = T^{\tau\mu} - \pi_\perp^{\tau\mu}$ whose components read \cite{Bazow:phdthesis}
\bs
\label{eq57}
\beal
   &K^\tau = (\ene{+}\Pperp) (u^\tau)^2 - \Pperp + \frac{\Delta\mathcal{P}(\tau u^\eta)^2}{1{+}u_\perp^2}
                 + \frac{2 W_{\perp z}^\tau \tau u^\eta}{\sqrt{1{+}u_\perp^2}}, \\
   &K^i = (\ene{+}\Pperp) u^\tau u^i  +   \frac{W_{\perp z}^i \tau u^\eta}{\sqrt{1{+}u_\perp^2}} ,
              \quad\qquad (i = x,y) \\
   &K^\eta = (\ene{+}\Pperp) u^\tau u^\eta + \frac{ \Delta \mathcal{P} u^\tau u^\eta}{1{+}u_\perp^2} 
                 + W_{\perp z}^\tau \frac{ (u^\tau)^2 + (\tau u^\eta)^2}{\tau u^\tau \sqrt{1{+}u_\perp^2}}. 
\end{align}
\es
In the last equation we used the orthogonality relation $z_\mu \Wperp = 0$ to eliminate $W_{\perp z}^\eta$. Taking the combination $\bigl(u^\eta\big)^2 K^\tau - u^\tau u^\eta K^\eta$ one obtains
\be
\label{eq58}
  \frac{u^\eta}{u^\tau} \equiv F = \frac{A-B\sqrt{1{-}\tilde{A}^2{+}\tilde{B}^2}}{1 + \tilde{B}^2},
\ee
where
\bs
\label{eq59}
\beal
  &A = \frac{K^\eta}{K^\tau + \PL}, \\
  &B = \frac{W^\tau_{\perp z}}{\tau\left(K^\tau + \PL\right)},
\end{align}
\es
as well as $\tilde{A} = \tau A$ and $\tilde{B} = \tau B$ are all known quantities. One can further use the normalization condition $u_\mu u^\mu{\,=\,}1$ to rewrite Eq.~\eqref{eq58} either as
\be
\label{eq60}
   \frac{u^\eta}{\sqrt{1{+}u_\perp^2}} = \frac{\tilde{F}}{\tau x}
\ee
or as
\be
\label{eq61}
  \frac{u^\tau}{\sqrt{1{+}u_\perp^2}} = \frac{1}{x},
\ee
where $\tilde{F} = \tau F$ and $x = \sqrt{1{-}\tilde{F}^2}$. With this the components of $z^\mu$ in~\eqref{eq55b} and thus $2W^{(\tau}_{\perp z} z^{\mu)}$ are now known. 

We next define the known vector $M^\mu{\,=\,}K^\mu{-}2W^{(\tau}_{\perp z} z^{\mu)}$, with components
\bs
\label{eq62}
\beal
   &\label{eq62a} M^\tau = (\ene{+}\Pperp) u^\tau u^\tau - \Pperp + \Delta \mathcal{P} (\tilde{F} / x)^2, \\
   &\label{eq62b} M^i = (\ene{+}\Pperp) u^\tau u^i, \quad\qquad (i = x,y) \\
   &\label{eq62c} M^\eta = (\ene{+}\Pperp) u^\tau u^\eta  + \Delta \mathcal{P} \tilde{F} / (\tau x^2).
\end{align}
\es
From Eq.~\eqref{eq62b} one obtains immediately the transverse flow velocity components and magnitude:
\bs
\label{eq63}
\beal
   &\label{eq63a} u^i = \frac{M^i}{u^\tau (\ene + \Pperp)}, \\
   &\label{eq63b} u_\perp = \frac{M_\perp}{u^\tau (\ene + \Pperp)} ,
\end{align}
\es
where $M_\perp = \sqrt{(M^x)^2{+}(M^y)^2}$. The two remaining unknown variables are $u^\tau$ and $\ene$. By taking the combination $ u_\perp^2 M^\tau - u^\tau u^i M^i$ one finds a relation between $u^\tau$ and $\ene$:
\be
\label{eq64}
   u^\tau = \sqrt{\frac{M^\tau + \Pperp - \Delta \mathcal{P} (\tilde{F} /x)^2}{\ene + \Pperp}}.
\ee
This can now be used to express $u^i$ and $u_\perp$ in Eqs.~(\ref{eq63}) as well as $u^\eta$ in Eq.~(\ref{eq58}) in terms of $\ene$ and the other known quantities. With a bit of algebra, and making use of the relation $\tilde{F} = \tau M^\eta / (M^\tau{+}\PL)$, the normalization condition $(u^\tau)^2 - u_\perp^2 - (\tau u^\eta)^2 = 1$ then yields the following explicit reconstruction formula for the energy density:
\be
\label{eq65}
\begin{split}
  \ene = &\ M^\tau - \Delta \mathcal{P}\, (\tilde{F} / x)^2 
                            - \frac{M_\perp^2}{M^\tau + \Pperp - \Delta \mathcal{P}\, (\tilde{F} /x)^2 } \\
             &\qquad  - \frac{(\tau M^\eta)^2 \big(M^\tau + \Pperp - \Delta \mathcal{P}\, (\tilde{F} /x)^2\big)}
                                      {(M^\tau + \PL)^2}.
\end{split}
\ee 
Note that, since $\PL$ and $\Pperp$ are evolved directly, the r.h.s. of Eq.~\eqref{eq65} is entirely known. This is in contrast with the reconstruction formula for $\ene$ in viscous hydrodynamics where one must solve some nonlinear equation $\mathcal{F}(\ene) = 0$ numerically \cite{Shen:2014vra,Bazow:phdthesis}. Once Eq.~\eqref{eq65} is evaluated, the fluid velocity components $u^\tau$, $u^i$, and $u^\eta$ can be determined using Eqs.~\eqref{eq64}, \eqref{eq63} and (\ref{eq58}) consecutively.  

%%%%%%%%%%%%%%%%%%%%%%%%%%%%%%%%%%%%%%%%%%%%%%%%%%%%%%%%%
\subsection{Reconstructing the anisotropic parameters}
\label{sec4c}
%%%%%%%%%%%%%%%%%%%%%%%%%%%%%%%%%%%%%%%%%%%%%%%%%%%%%%%%%

Having reconstructed the LRF energy density $\ene$, the left hand sides of equations (\ref{eq53}) are all known, as is the temperature $T$ (from $\ene(T)$) and thus the particle mass $m(T)$. We can now use the expressions on the right hand sides of these equations to determine the anisotropic parameters $(\Lambda,\alpha_\perp,\alpha_L)$. We write equations (\ref{eq53}) as
\be
\label{eq66}
   F(X) = 0
\ee
where 
\be
\label{eq67}
   X = \left(
    \begin{array}{c}
   \Lambda \\ \alpha_\perp \\ \alpha_L 
   \end{array}
   \right)
\ee
and 
\be
\label{eq68}
   F(X) = \left(
   \begin{array}{c}
   \I_{2000}(X){-}\ene{+}B \\ \I_{2200}(X){-}\PL{-}B \\ \I_{2010}(X){-}\Pperp{-}B 
   \end{array}
   \right).
\ee
We then use Newton's method in three dimensions to find the solution vector $X$. One starts with an initial guess vector $X^{(0)}$, which is typically the value of $X$ at the spatial grid point known from the previous time step. This vector is updated with an iteration $\Delta X$, which is given by the matrix equation 
\be
\label{eq69}
   J_{ij} \, \Delta X_j = - F_i
\ee
where $J_{ij} = \partial F_i / \partial X_j$ is the Jacobian matrix. The analytical form of this matrix is
\be
\label{eq70}
   J = \left(
   \begin{array}{c c c}
       \dfrac{\J_{2001}}{\Lambda^2} \, & \,\dfrac{2\J_{401-1}}{\Lambda \alpha_\perp^3} \, & \, \dfrac{\J_{420-1}}{\Lambda \alpha_L^3} \\ \\
              \dfrac{\J_{2201}}{\Lambda^2}\, & \, \dfrac{2\J_{421-1}}{\Lambda \alpha_\perp^3} \, 
                                                                          & \, \dfrac{\J_{440-1}}{\Lambda \alpha_L^3} \\ \\
        \dfrac{\J_{2011}}{\Lambda^2} \, & \, \dfrac{4\J_{402-1}}{\Lambda \alpha_\perp^3} \, 
                                                                                  & \, \dfrac{\J_{421-1}}{\Lambda \alpha_L^3} \\
   \end{array}
   \right).
\ee
One can simplify the computation of the matrix elements in~\eqref{eq70} by using the identities (see Appendix~\ref{appe})
\bs
\label{eq71}
\beal
   &\J_{420-1} = \Lambda \alpha_L^2 (\I_{2000}+\I_{2200}), \\
   &\J_{401-1} = \Lambda \alpha_\perp^2 (\I_{2000}+\I_{2010}), \\
   &\J_{421-1} = \frac{\Lambda \alpha_\perp^2 \alpha_L^2}{\alpha_L^2{-}\alpha_\perp^2} (\I_{2200}-\I_{2010}).
\end{align}
\es
The iteration process~\eqref{eq69} is repeated until convergence is achieved. Newton's method is a local root-finder and works well if the initial guess $X^{(0)}$ is sufficiently close to the solution. This may not be the case when the hydrodynamic variables are first initialized or evolving too rapidly. To better handle such situations, we include a line-backtracking algorithm, which takes partial steps of $\Delta X$, to improve global convergence \cite{Press:1992:NRC:148286}.

Each iteration of Newton's method requires the numerical computation of eight one-dimensional integrals, three for $F$ and five for $J$. Alternatively, one may use Broyden's method, which approximates the Jacobian in terms of $F$, so only three integrals need to be evaluated. \cite{Press:1992:NRC:148286} An approximate Jacobian may complicate the line-backtracking search, which ensures a decrease of $\abs{F}$ only if the exact Jacobian is used. For iterations where the full Broyden step does not sufficiently decrease $\abs{F}$ we switch to Newton's method. 

%%%%%%%%%%%%%%%%%%%%%%%%%%%%%%%%%%%%%%%%%%%%%%%%%%%
\begin{figure*}[t]
\centering
\includegraphics[]{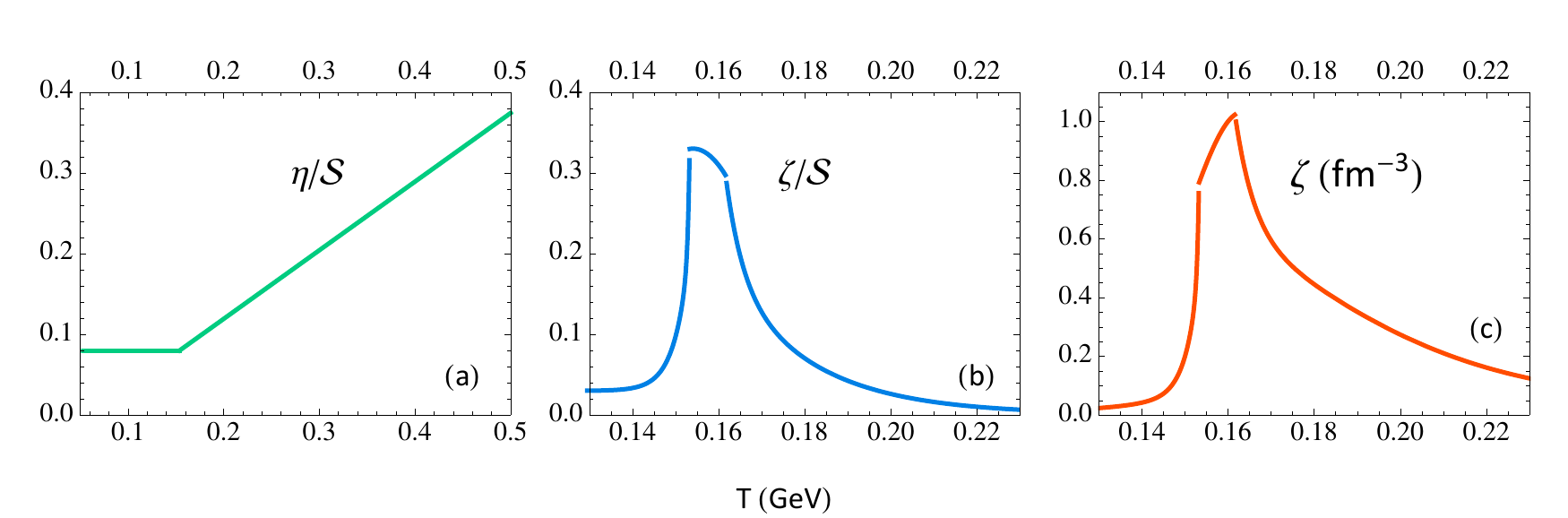}
\caption{(Color online)
	The temperature dependence of (a) the specific shear viscosity $\eta / \mathcal{S}$, (b) the specific 
	bulk viscosity $\zeta / \mathcal{S}$, and (c) the bulk viscosity $\zeta$ itself in units of fm$^{-3}$, given
	by the parameterizations (\ref{eq78}) and (\ref{eq79}) \cite{Bernhard:2016tnd}. We note that the peak 
	of $\zeta / \mathcal{S}$ in (b) occurs at the temperature $T = 0.995 \, T_c \approx T_c$ whereas the 
	bulk viscosity $\zeta = (\zeta / \mathcal{S}) \times \mathcal{S}$ in panel (c) peaks at the higher 
	temperature $T = 1.05 \, T_c$. 
\label{FV}
}
\end{figure*}
%%%%%%%%%%%%%%%%%%%%%%%%%%%%%%%%%%%%%%%%%%%%%%%%%%%

In the anisotropic equation of state model (\ref{eq53}) there is an ambiguity between the initialization of the mean field $B$ and of the kinetic terms $\ene^{(k)}$, $\PL^{(k)}$, and $\Pperp^{(k)}$. Standard hydrodynamic initial conditions for the energy-momentum tensor only provide $\ene$, $\PL$, and $\Pperp$ on the initialization hypersurface. The initial energy density profile $\ene$ also yields the initial temperature profile and thus the initial profile for the equilibrium part $B_\eq(T)$ of the mean field. Its comoving time derivative on the initialization surface can be obtained by taking the equilibrium limit of Eq.~(\ref{eq52}): 
\be
\label{eq72}
   \dot{B}_\eq = -\frac{\dot m}{m}\left( \ene_\eq^{(k)}{-}3\Peq^{(k)}\right). 
\ee
To obtain a guess for the initial non-equilibrium deviation $\delta B$ we assume that $\delta B$ evolves on a time scale larger than the bulk viscous relaxation time $\tau_\Pi$. We can then ignore the time derivative of $\delta B$ on the left hand side of Eq.~(\ref{eq52}) and obtain from the difference between Eqs.~(\ref{eq52}) and (\ref{eq72}) the ``asymptotic'' initial condition
\be
\label{eq73}%eq75
   \delta B^{\mathrm{(asy)}} = \frac{3 \tau_\Pi \dot m}{m - 4 \tau_\Pi \dot m}\Pi.
\ee
As before (see footnote \ref{fn5}) $m$ and $\dot{m}$ can be expressed in terms of the energy density $\ene$ and its LRF time derivative. Having thus specified the initial profile for the mean field $B = B_\eq + \delta B$ we can proceed to extract the initial anisotropic parameters and compute the initial values for the transport coefficients. 

For far-from-equilibrium initial conditions, such as those provided by the IP-Glasma model \cite{Schenke:2012wb} where $\PL$ starts out with a very large negative value $\PL=-\ene$ and, after classical Yang-Mills evolution for a time of the order of the inverse saturation momentum, settles to around zero \cite{Gelis:2013rba,Wang:2034451}, the implied deviation $\PL{-}\Peq$ can become so large that, with this initial choice of $B$, the quasiparticle model cannot accommodate it within the allowed ranges for $(\Lambda,\alpha_\perp,\alpha_L)$. Since typically $B{\,<\,}0$ at high temperatures (see Fig.~\ref{F1}), the kinetic longitudinal pressure, $\PL^{(k)} = \PL{+}B$, may in this situation be negative. Specifically, the anisotropic parameter initialization is found to fail when $\PL/\Pperp \lesssim 0.08$. To overcome this problem, in the case of such extreme initial conditions for $\PL$ we simply adjust our initial guess for $\delta B$ and increase the initial value for $B$ until a solution for $(\Lambda,\alpha_\perp,\alpha_L)$ can be found. More meaningful ways of dealing with this shortcoming will be left to future work. 

%%%%%%%%%%%%%%%%%%%%%%%%%%%%%%%%%%%%%%%%%%%%%%%%%%%%%%%
\section{Bjorken Flow}
\label{sec5}
%%%%%%%%%%%%%%%%%%%%%%%%%%%%%%%%%%%%%%%%%%%%%%%%%%%%%%%

In this Section we test our anisotropic hydrodynamic formalism by comparing it to standard viscous hydrodynamics  for the case of (0+1)-dimensional Bjorken expansion, using the state-of-the-art lattice QCD equation of state referenced in Fig.~\ref{F1}. We begin by simplifying the anisotropic evolution equations (\ref{eq10})-(\ref{eq12}), (\ref{eq35})-(\ref{eq36}), and (\ref{eq54})-(\ref{eq55}) for systems with Bjorken symmetry. In Milne coordinates, the fluid velocity is $u^\mu = (1,0,0,0)$, the longitudinal and transverse expansion rates are $z_\mu D_z u^\mu = 1/\tau$ and $\theta_\perp = 0$, and the residual shear stress components $\Wperp$ and $\piperp$ vanish by symmetry. The component $T^{\tau\tau}$ trivially reduces to $\ene$. As a result, the anisotropic hydrodynamic equations simplify to
\bs
\label{eq:ahydroeqs}%eq76 
\beal
   &\dot\ene =  - \frac{\ene+\PL}{\tau},
   \\
   &\dot{\mathcal{P}}_L =  - \frac{\bar{\mathcal{P}}{-}\Peq}{\tau_\Pi} 
      - \frac{\PL{-}\Pperp}{3\tau_\pi / 2} + \frac{\bar{\zeta}^L_z}{\tau},
   \\
   &\dot{\mathcal{P}}_\perp =  - \frac{\bar{\mathcal{P}}{-}\Peq}{\tau_\Pi}
     + \frac{\PL{-}\Pperp}{3\tau_\pi} + \frac{\bar{\zeta}^\perp_z}{\tau},
   \\
   &\dot{B} = -\frac{B{-}B_\eq}{\tau_\Pi} 
   + \frac{\ene{+}\PL}{\tau m}\frac{dm}{dT}\frac{dT}{d\ene}\big(\ene{-}2\Pperp{-}\PL{-}4B \big),
\end{align}
\es
where in (\ref{eq:ahydroeqs}d) we used
\be
  \label{eq75}%eq77
   \frac{\dot{m}}{m} = -\frac{\ene{+}\PL}{\tau m}\frac{dm}{dT}\frac{dT}{d\ene}
\ee
as well as $\ene^{(k)}{-}2\Pperp^{(k)}{-}\PL^{(k)} = \ene{-}2\Pperp{-}\PL{-}4B$. 

To fix the relaxation times $\tau_\pi$ and $\tau_\Pi$ we proceed as follows: we use a temperature dependent parametrization for the specific shear viscosity $\eta/\mathcal{S}$ \cite{Bernhard:2016tnd},
\be
\label{eq78}%eq78
\eta/\mathcal{S} = \bigg\{
\begin{array}{ll}
      (\eta/\mathcal{S})_{\text{min}} + (\eta/\mathcal{S})_{\text{slope}}(T{-}T_c) & \text{for} \ T > T_c\,,
  \\
      (\eta/\mathcal{S})_{\text{min}} & \text{for} \ T\leq T_c \,,
\end{array} 
\ee
where $\mathcal S = \mathcal S(\ene)$ is the lattice QCD entropy density and $T_c = 154$\,MeV is the pseudo-critical temperature. The model parameters $(\eta/\mathcal S)_{\text{min}} = 0.08$ and $(\eta/\mathcal S)_{\text{slope}} = 0.85$\,GeV$^{-1}$ were extracted from a global Bayesian analysis of RHIC and LHC heavy-ion collision data \cite{Bernhard:2016tnd}.\footnote{%
	Note that, while some of these parameters where fitted to experimental data \cite{Bernhard:2016tnd}, 
	this was done with standard viscous hydrodynamics, and slightly different values might be expected when 
	repeating that exercise with anisotropic hydrodynamics. The precise values of the parameters in 
	Eqs.~(\ref{eq78})-(\ref{eq80}) should therefore not be taken too seriously.
	}
Similarly, we use for the specific bulk viscosity $\zeta/\mathcal S$ the parameterization from Ref.~\cite{Denicol:2009am}: 
\be
\label{eq79}%eq79
   \zeta/\mathcal{S} = (\zeta/\mathcal{S})_{\text{norm}}   f(T/T_p)
\ee
where the function $f(x)$ is  given by
\be
\label{eq80}%eq80
f = \Bigg\{
\begin{array}{ll}
      C_1 + \lambda_1 \exp\big[\frac{x-1}{\sigma_1}\big] + \lambda_2\exp\big[\frac{x-1}{\sigma_2}\big] 
      & (x{\,<\,}0.995), \\
      A_0 + A_1 x + A_2 x^2 & \hspace*{-1cm} (0.995{\,\leq\,}x{\,\leq\,}1.05), \\
      C_2 + \lambda_3 \exp\big[\frac{1-x}{\sigma_3}\big] + \lambda_4\exp\big[\frac{1-x}{\sigma_4}\big]  & (x{\,>\,}1.05),
\end{array} 
\ee
with $A_0 = -13.45$, $A_1 = 27.55$, $A_2 = -13.77$, $C_1 = 0.03$, $C_2 = 0.001$, $\lambda_1 = 0.9$, $\lambda_2 = 0.22$, $\lambda_3 = 0.9$, $\lambda_4 = 0.25$, $\sigma_1 = 0.0025$, $\sigma_2 = 0.022$, $\sigma_3 = 0.025$ and $\sigma_4 = 0.13$. For the normalization factor we choose $(\zeta / \mathcal S)_{\text{norm}} = 1.25$ \cite{Bernhard:2016tnd}, and we fix the location of the peak of the specific bulk viscosity by taking $T_p = T_c$.\footnote{%
	Note that this puts the peak of the bulk viscosity at a much lower temperature than assumed in most previous
	implementations of this parametrization (see e.g. \cite{Denicol:2009am, Ryu:2015vwa, Paquet:2015lta, 
	Bernhard:2016tnd, Vujanovic:2017fkh}).
	}  
Fig.~\ref{FV} shows the behavior of the specific shear and bulk viscosities as a function of temperature. 

%%%%%%%%%%%%%%%%%%%%%% Fig. 2 %%%%%%%%%%%%%%%%%%%%%%%%%%%%%%
\begin{figure*}[t]
\centering
\includegraphics[width=\textwidth]{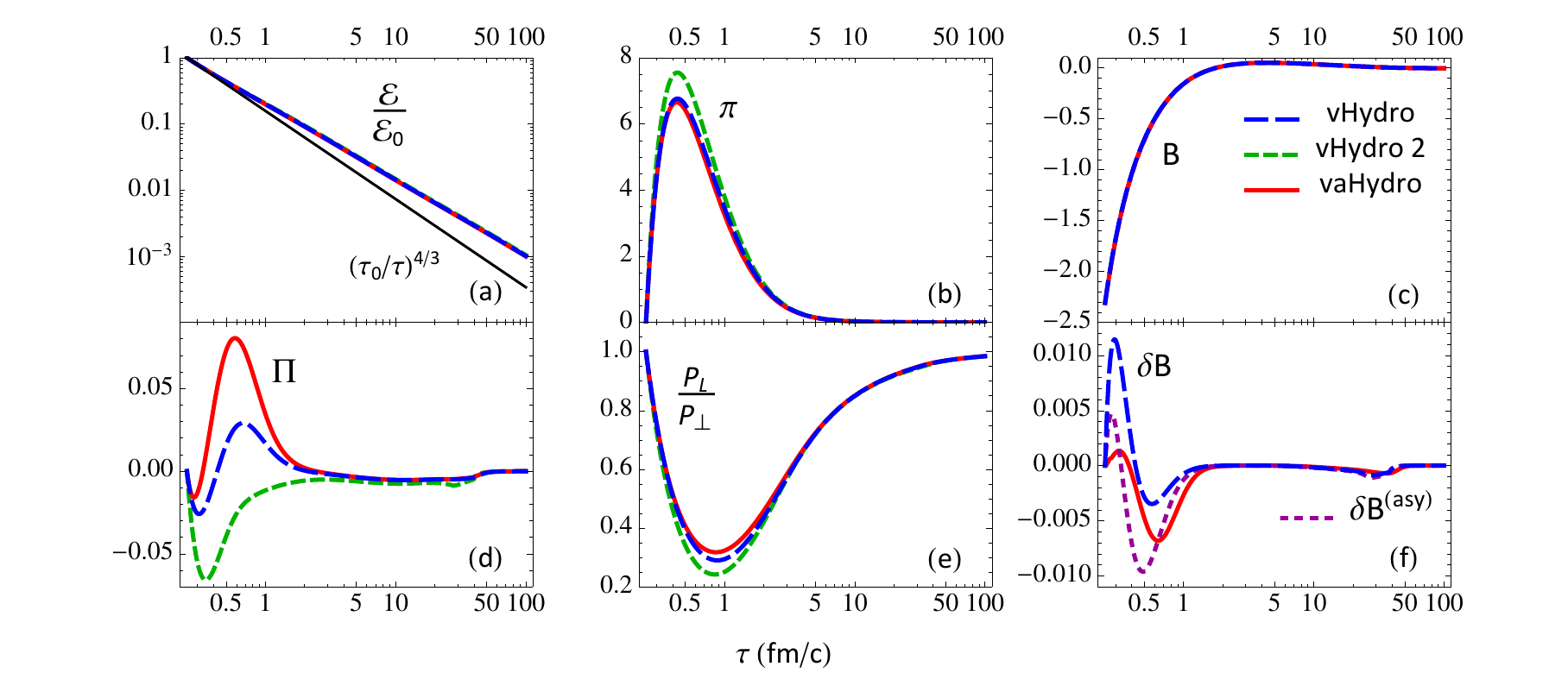}
\caption{(Color online) 
	The Bjorken evolution of the normalized energy density $\ene/\ene_0$, bulk viscous 
	pressure $\Pi$, longitudinal shear stress component $\pi$, pressure anisotropy $\PL/\Pperp$,
	total mean field $B$, and non-equilibrium component of the mean field $\delta B$, calculated in 
	anisotropic hydrodynamics (solid red, vaHydro) as well as in standard viscous hydrodynamics 
	(dashed blue and green, vHydro and vHydro\,2). The solid red and long-dashed blue lines (vaHydro 
	and vHydro) use transport coefficients derived from kinetic theory for medium-dependent 
	quasiparticles while the dashed green lines (vHydro 2) use kinetic theory transport coefficients 
	derived in the small fixed mass limit. The fluid starts out in thermal equilibrium at longitudinal 
	proper time $\tau_0 = 0.25$\,fm/$c$ with initial temperature $T_0 = 0.5$\, GeV. In panel (f), 
	the short-dashed purple line shows the ``asymptotic approximation'' \eqref{eq73} for $\delta B$, 
	computed using data from the anisotropic hydrodynamic evolution whereas the long-dashed blue
	line uses Eq.~(\ref{eq:dB2nd}) and data from the viscous hydrodynamic evolution. $\Pi$, $\pi$, $B$, 
	and $\delta B$ are plotted in units of GeV/fm$^3$.
	\label{F2}}
\end{figure*}
%%%%%%%%%%%%%%%%%%%%%%%%%%%%%%%%%%%%%%%%%%%%%%%%%%%%%%%
%

The relaxation times are then obtained from the kinetic theory relations (\ref{eq38}), rewritten in the form
\be
\label{eq81}%eq81
   \tau_\pi = \frac{\eta}{\mathcal{S}} \, \frac{\mathcal{S}}{\beta_\pi} , \qquad
   \tau_\Pi = \frac{\zeta}{\mathcal{S}} \, \frac{\mathcal{S}}{\beta_\Pi},
\ee
using the following quasiparticle versions of the isotropic thermodynamic integrals $\beta_\pi$ and $\beta_\Pi$ \cite{Tinti:2016bav}:
\begin{eqnarray}
\label{eq:beta}%eq82
    && \beta_\pi = \frac{1}{T} \int_p \frac{(- p \cdot \Delta \cdot p)^2}{15(\up)} f_\eq ,
  \\\nonumber
    && \beta_\Pi = \frac{5\beta_\pi}{3} - c_s^2(\ene{+}\Peq) 
       + c_s^2 m \frac{dm}{dT} \int_p \frac{- p \cdot \Delta \cdot p}{3(\up)} f_\eq .
\end{eqnarray}
Here $c_s^2(\ene)$ is the squared speed of sound from lattice QCD. The system of ordinary differential equations \eqref{eq:ahydroeqs} is solved using Huen's method. After each intermediate and full time step the anisotropic parameters are updated by numerically solving Eq.~\eqref{eq66}.

These anisotropic hydrodynamic results will be compared with those from second-order viscous hydrodynamics in the 14-moment approximation. The corresponding evolution equations and transport coefficients are derived in Appendix~\ref{appd}. For Bjorken flow, the set of independent dynamical variables reduces to the energy density $\ene$, the shear stress $\pi = -\tau^2 \pi^{\eta\eta} = \frac{2}{3}(\Pperp{-}\PL)$, and the bulk viscous pressure $\Pi = \frac{1}{3}(2\Pperp{+}\PL) - \Peq$. Their evolution equations simplify to
\bs
\label{eq:vhydroeqs}%eq83
\beal
   & \dot\ene=  - \frac{\ene+\Peq + \Pi - \pi}{\tau} ,
   \\
   & \dot\pi = - \frac{\pi}{\tau_\pi} - \frac{4 \beta_\pi}{3\tau} - \frac{\Big(\tau_{\pi\pi}
       +3\delta_{\pi\pi}\Big)\pi - 2\lambda_{\pi\Pi}\Pi}{3\tau_\pi \tau},
   \\
   & \dot\Pi =  - \frac{\Pi}{\tau_\Pi} - \frac{\beta_\Pi}{\tau} - \frac{\delta_{\Pi\Pi}\Pi - \lambda_{\Pi\pi}\pi}{\tau_\Pi \tau}.
\end{align}
\es
For the non-equilibrium mean field contribution $\delta B$ we use the second-order expression \cite{Tinti:2016bav}
\be
\label{eq:dB2nd}%eq84
  \delta B^{(2)} = - \frac{3 \tau_\Pi}{m} \frac{dm}{dT} \frac{dT}{d\ene} (\ene+\Peq) \, \Pi \, \theta 
\ee
where $\theta{\,=\,}\del_\mu u^\mu{\,=\,}1/\tau$ is the scalar expansion rate. In Eq.~\eqref{eq:dB2nd}, we replaced the relaxation time $\tau_r$ by $\tau_\Pi$. The relaxation times $\tau_\pi$ and $\tau_\Pi$ are obtained from Eqs.~(\ref{eq81}) and \eqref{eq:beta} while the second-order transport coefficients $\tau_{\pi\pi}, \, \delta_{\pi\pi}, \, \lambda_{\pi\Pi}, \, \delta_{\Pi\Pi}$ and $\lambda_{\Pi\pi}$ are computed from the quasiparticle model in the 14-moment approximation (after expansion around a local equilibrium distribution, see Appendix~\ref{appd}). We will also look at how the transport coefficients, including the relaxation times, affect the viscous hydrodynamic results when computed in the small fixed mass approximation $z \ll 1$ and $dm/dT \approx 0$, without a mean field, which is commonly implemented in viscous hydrodynamic simulations. \cite{Denicol:2014vaa,Ryu:2015vwa,Ryu:2017qzn}

%%%%%%%%%%%%%%%%%%%%%%%%%%%%%%%%%%%%%%%%%%%%%%%%%%%%%%%
\subsection{Equilibrium initial conditions}
\label{sec5a}
%%%%%%%%%%%%%%%%%%%%%%%%%%%%%%%%%%%%%%%%%%%%%%%%%%%%%%%

%%%%%%%%%%%%%%%%%%% Fig. 3 %%%%%%%%%%%%%%%%%%%%%%%%%%%%%%%%% 
\begin{figure*}[t]
\centering
\includegraphics[]{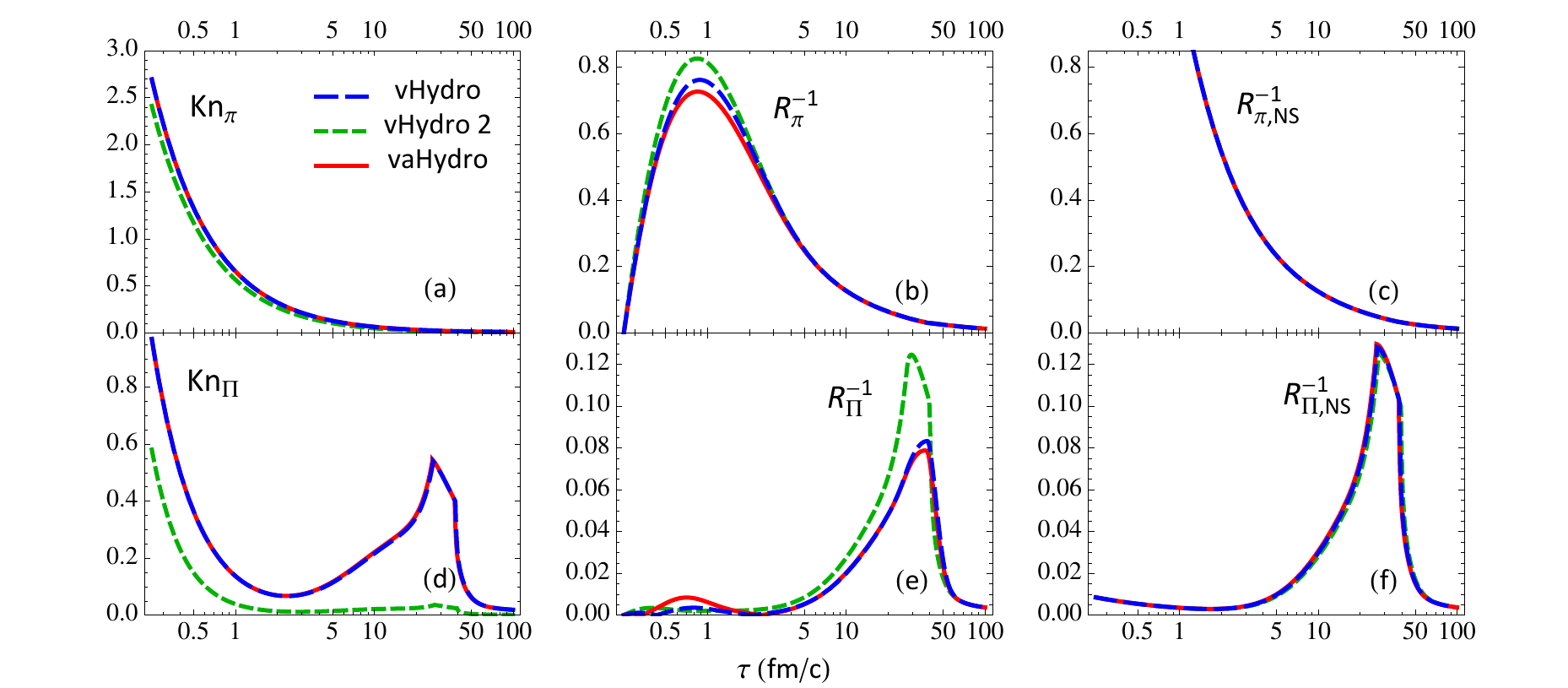}
\caption{(Color online)
	The shear (top row) and bulk (bottom row) Knudsen and inverse Reynolds numbers numbers 
	associated with Fig.~\ref{F2}. For Bjorken flow, the formulas for the Knudsen and inverse Reynolds  
	numbers reduce to $\text{Kn}_\pi = \tau_\pi \sqrt{\sigma_\munu \sigma^\munu} = 
	\sqrt{2/3} \, \tau_\pi / \tau$, $\text{Kn}_\Pi = \tau_\Pi \theta =  \tau_\Pi / \tau$, 
	$R^{-1}_\pi = \sqrt{\pi_\munu \pi^\munu} /\Peq =  \sqrt{3/2} \, \pi / \Peq$, and $R^{-1}_\Pi = 
	\abs{\Pi} / \Peq$. The last column (panels (c,f)) shows the Navier-Stokes limits of the shear and 
	bulk inverse Reynolds numbers, $R^{-1}_{\pi,\mathrm{NS}} = \sqrt{8/3}\,\eta/(\tau\Peq)$ and 
	$R^{-1}_{\Pi,\mathrm{NS}} = \zeta/(\tau\Peq)$.
\label{F3}
}
\end{figure*}
%%%%%%%%%%%%%%%%%%%%%%%%%%%%%%%%%%%%%%%%%%%%%%%%%%%%%%%

In Figure~\ref{F2} we show the Bjorken evolution of the hydrodynamic variables in anisotropic hydrodynamics, including the total mean field and its non-equilibrium component in the quasiparticle (QP) model used to compute the transport coefficients, and compare it with that in the standard viscous hydrodynamic models. Figure~\ref{F3} shows the same for the associated Knudsen and inverse Reynolds numbers. In this subsection we impose equilibrium initial conditions with initial temperature $T_0=0.5$\,GeV at longitudinal proper time $\tau_0=0.25$\,fm/$c$, i.e. all non-equilibrium effects are initially zero. Figure~\ref{F2}a shows that all three models (anisotropic hydrodynamics with QP transport coefficients in solid red lines, standard viscous hydrodynamics with QP transport coefficients in long-dashed blue lines and transport coefficients from a Boltzmann gas in a small fixed mass expansion in short-dashed green lines) produce almost identical evolutions for the energy density. The energy density decreases somewhat more slowly than for a conformal ideal fluid, indicated by the thin black line ${\sim\,}\tau^{-4/3}$. This is due to the smaller pressure of our EOS (which thus performs less longitudinal work) and to viscous heating. For reference we note that the system passes through the pseudo-critical temperature $T_c{\,=\,}154$\,MeV at $\tau_c{\,\sim\,}37$\,fm/$c$, with a small spread of less than 2\,fm/$c$ between the three models. 

Panel (c) shows that, if a QP model is used for the transport coefficients, the mean field $B$ also evolves almost identically in anisotropic and standard viscous hydrodynamics. Small differences between anisotropic and standard viscous hydrodynamics with QP transport coefficients are observed in the evolution of the shear stress $\pi$ ($\order{(2\%)}$) and the pressure ratio $\PL/\Pperp$ ($\order{(10\%)}$): the effective resummation of shear viscous effects in anisotropic hydrodynamics leads to a slight reduction of the shear stress, resulting in a slightly reduced pressure anisotropy. Standard viscous hydrodynamics, with transport coefficients calculated in the small fixed mass expansion (short-dashed green lines), produces somewhat ($\order{(15\%)}$) larger shear stresses and stronger pressure anisotropies.

Given that the pressure anisotropy gets quite large, with $\PL/\Pperp$ decreasing to about 30\% at $\tau{\,\sim\,}1$\,fm/$c$, the excellent agreement between standard viscous and anisotropic hydrodynamics is somewhat unexpected. It suggests that the widely used standard viscous hydrodynamic approach is quite robust and quantitatively reliable even for large shear stresses. Similar observations were made before in \cite{Bazow:phdthesis} as well as in studies of the Bjorken dynamics of strongly coupled theories where second-order viscous hydrodynamics could be directly compared with an exact numerical solution of the underlying strong-coupling dynamics \cite{Chesler:2015lsa, Chesler:2015bba}.   

%%%%%%%%%%%%%%%%%%% Fig. 4 %%%%%%%%%%%%%%%%%%%%%%%%%%%%%%%%%
\begin{figure*}[t]
\centering
\includegraphics[width=\textwidth]{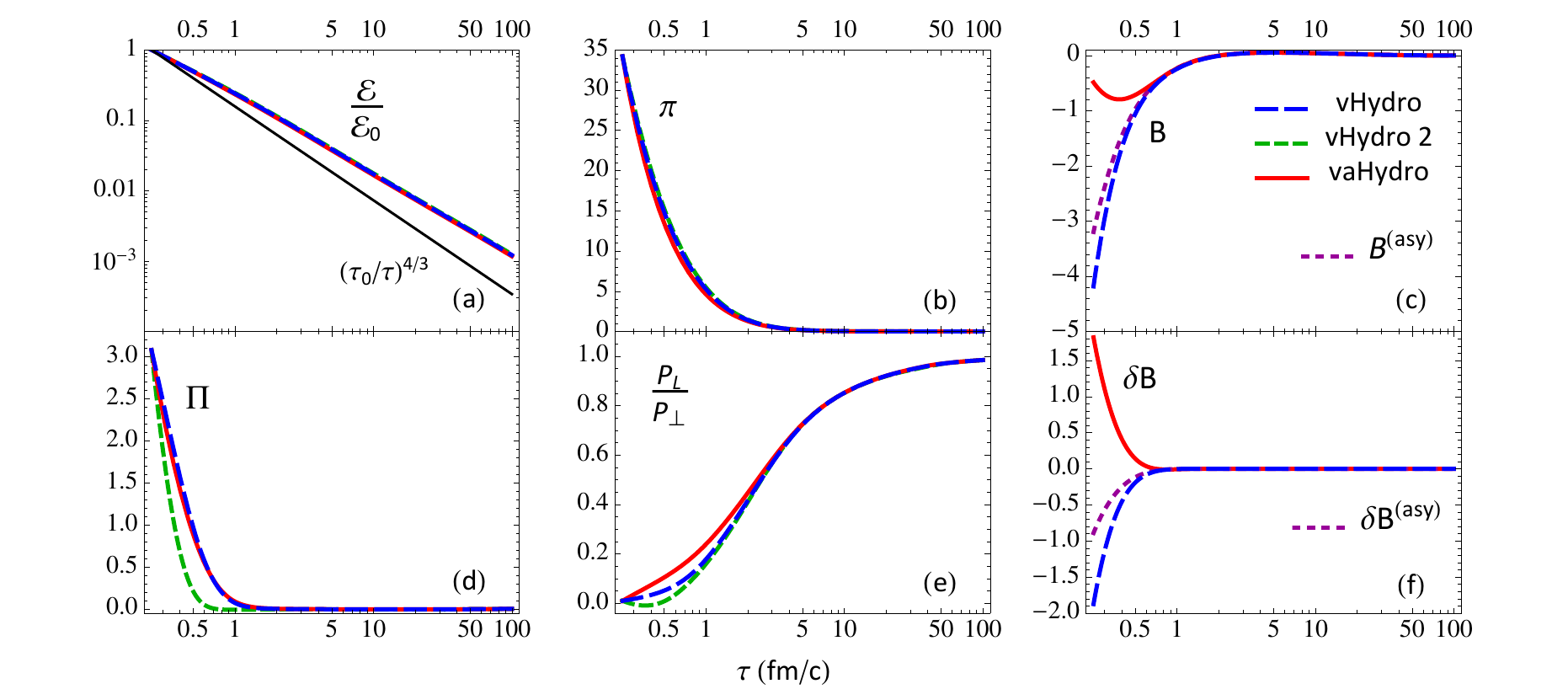}
\caption{(Color online)
	Same as Fig.~\ref{F2} but for Glasma-like initial conditions, with the same initial temperature at the 
	same initial time. The initial pressures are set to $\mathcal{P}_{L0} = 4.975 \times 10^{-3} \, \ene_0$ 
	and $\mathcal{P}_{\perp0} = 0.4975 \, \ene_0$. For the anisotropic hydrodynamic evolution (solid red
	line) the magnitude of the initial mean field $B_0$ is reduced to $15.3\,\%$ of the default value 
	(purple short-dashed line) $B^{\mathrm{(asy)}} = B_\eq + \delta B^{\mathrm{(asy)}}$. For the standard 
	viscous hydrodynamic evolution with quasiparticle transport coefficients (long-dashed blue lines) $B$
	and $\delta B$ are determined as described in the text. $\Pi$, $\pi$, $B$, and $\delta B$ are plotted 
	in units of GeV/fm$^3$. 
\label{F4}
}
\end{figure*}
%%%%%%%%%%%%%%%%%%%%%%%%%%%%%%%%%%%%%%%%%%%%%%%%%%%%%%%

The largest differences between anisotropic and standard viscous hydrodynamics are seen in the evolution of the bulk viscous pressure $\Pi$ (Fig.~\ref{F2}d) and the non-equilibrium part of the mean field $\delta B$ (Fig.~\ref{F2}f). The two panels expose strong correlations between the evolutions of these two quantities. Both are small: (i) The bulk viscous pressure at early times is about 100 times smaller than the shear stress. While the evolution of $\Pi$ is qualitatively similar (although quantitatively different by more than a factor 2 at early times) for anisotropic and standard viscous hydrodynamics with QP transport coefficients, it exhibits {\em qualitatively different} dynamics in standard viscous hydrodynamics with transport coefficients computed from the small fixed mass expansion. (ii) Compared to the equilibrium mean field, the non-equilibrium part $\delta B$ is about two orders of magnitude smaller (see panels (c) and (f) of Fig.~\ref{F2}). Here one observes very different trajectories for $\delta B$ between the evolutions from anisotropic and standard viscous hydrodynamics, although their shapes are qualitatively similar. In addition, panel (f) shows for comparison the ``asymptotic approximation'' (\ref{eq73}) (short-dashed purple curve) which should be compared to the exact numerical solution (red solid line). Obviously, the large expansion rate at early times makes the asymptotic trajectory, which is based on the assumption that $\delta B$ evolves more slowly than the bulk relaxation rate, a rather crude approximation.

Figure~\ref{F3} shows the Knudsen and inverse Reynolds numbers associated with the shear and bulk viscous stresses. While the Knudsen and inverse Reynolds numbers associated with shear stress dominate the non-equilibrium dynamics at early times, those associated with bulk viscosity are the most relevant at late times when the system passes through the QCD phase transition. \footnote{%
	Note that the Knudsen numbers for vaHydro (solid red) and vHydro (long-dashed blue) are almost 
	identical due to the very similar energy density and temperature evolution, see Eq.~(\ref{eq:beta}) and 
	Fig.~\ref{F2}a.
	}

In spite of the shear Knudsen number (Fig.~\ref{F3}a) starting out large with a value of around 2.5, the shear inverse Reynolds number (Fig.~\ref{F3}b) never exceeds a value of about 75-85\%. This results from the delay caused by the microscopic shear relaxation time which controls the approach of the shear stress $\pi$ from its zero starting point to its Navier-Stokes value and has at $\tau_0=0.25$ the value $\tau_\pi\approx 0.8$\,fm/$c$. By the time $R_\pi^{-1}$ reaches its Navier-Stokes limit (shown in Fig.~\ref{F3}c), the shear Knudsen number has already dropped to values well below 1. We reiterate that at the peak value $\sim 3/4$ of the shear inverse Reynolds number the differences between anisotropic and standard viscous hydrodynamic evolution are less than 6\% as long as both are evaluated with transport coefficients computed from the same underlying kinetic theory.

\begin{figure*}[t]
\centering
\includegraphics[]{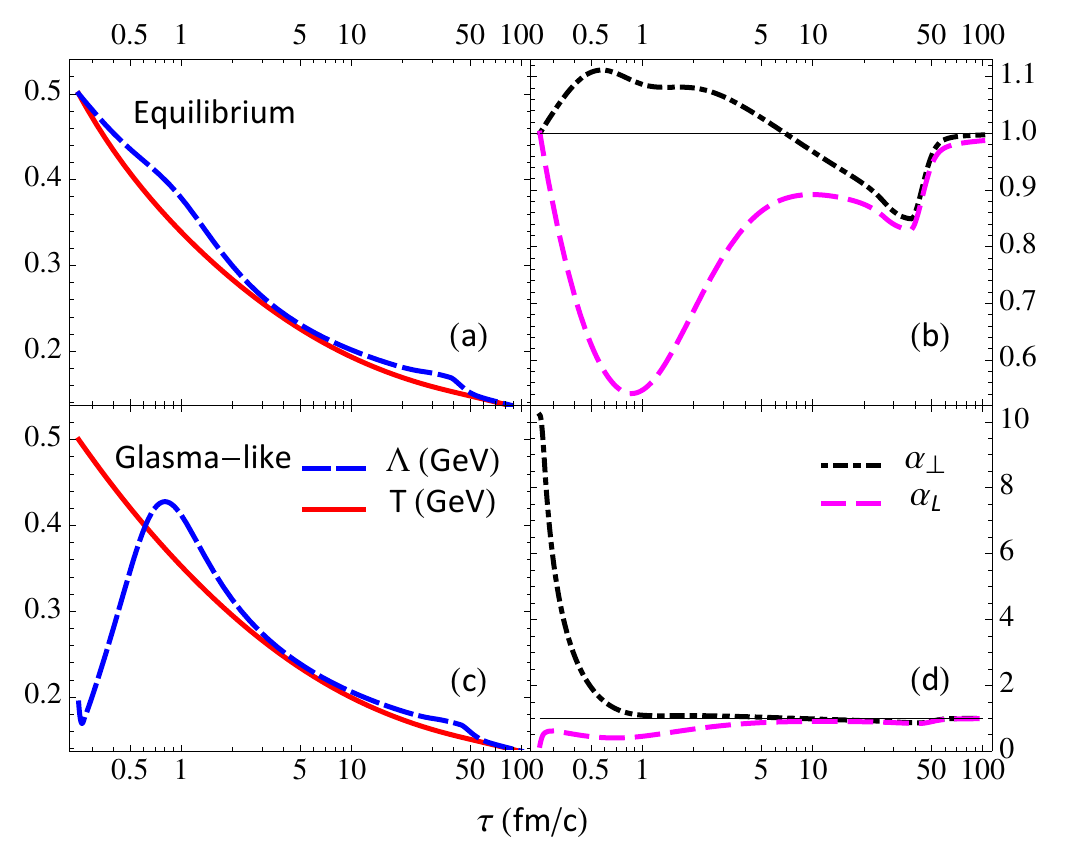}
\caption{(Color online)
	The anisotropic hydrodynamic evolution of the effective temperature $\Lambda$ (dashed blue) 
	compared to the temperature $T$ (solid red) is shown in the left panels for equilibrium (a) and 
	Glasma-like initial conditions (c). The right panels show the evolution of the momentum deformation 
	parameters $\alpha_L$ (dashed magenta) and $\alpha_\perp$ (dash-dotted black), again for 
	equilibrium (b) and Glasma-like initial conditions (d). The temperatures $\Lambda$ and $T$ are 
	plotted in units of GeV. The deformation parameters $\alpha_L$ and $\alpha_\perp$ are unitless. 
\label{F6}
}
\end{figure*}

Fig.\,\ref{F3}e shows the evolution of the bulk inverse Reynolds number $R_\Pi^{-1}$, which peaks due to critical dynamics near the QCD phase transition temperature $T_c$; the corresponding Navier-Stokes value is shown in Fig.~\ref{F3}f. Because $\tau_\Pi\propto\zeta$ (see Eq.~(\ref{eq81})), the bulk relaxation rate slows down when the bulk viscosity peaks. This leads to ``critical slowing down'' of the evolution of the bulk viscous pressure $\Pi$, limiting its growth as the system cools down to $T_c$ \cite{Berdnikov:1999ph, Song:2009rh}. Comparing the solid red and dashed blue curves in Figs.~\ref{F3}e and f we see that $\Pi$ and thus $R_\Pi^{-1}$ never reaches much more than about half of its peak Navier-Stokes value, and it also peaks later (around $\tau{\,\sim\,}38$\,fm/$c{\,\approx\,}\tau_c$, corresponding to $T{\,\approx\,}0.995\,T_c$) than the Navier-Stokes limit which reaches its maximum already at $\tau{\,\sim\,}27$\,fm/$c$ (corresponding to $T{\,\approx\,}1.05\,T_c$). One observes that even near its peak at $\tau \sim38$\,fm/$c$, $R_\Pi^{-1}$ evolves almost identically in anisotropic and standard viscous hydrodynamics with QP transport coefficients.\footnote{%
	The reason for this will become clearer in Fig.~\ref{F6} below where we will see that at late 
	times anisotropic hydrodynamics reduces in good approximation to 2nd order viscous hydrodynamics.
	} 

A marked difference is observed, however, when the system is evolved with standard hydrodynamics using transport coefficients from a massless Boltzmann gas without a mean field (green short-dashed lines in Figs.~\ref{F3}d,e). It turns out that the thermodynamic integral $\beta_\Pi$ in Eq.~(\ref{eq:beta}) is remarkably sensitive to the degree of nonconformality of the Boltzmann gas, giving rise to a much longer bulk viscous relaxation time in the QP model than for the light Boltzmann gas without a mean field, especially in the neighborhood of $T_c$. This is reflected in the large difference between the short-dashed green line and the other two curves for the bulk Knudsen number shown in Fig.~\ref{F3}d which causes the corresponding large difference in the evolution of the bulk inverse Reynolds number shown in panel (e): The much shorter relaxation time for the light Boltzmann gas allows the bulk viscous pressure to follow its Navier-Stokes limit (shown in Fig.~\ref{F3}f) much more closely, causing $R_\Pi^{-1}$ to rise much more steeply and to a larger peak value as the system cools towards $T_c$ than in the other two approaches where $\beta_\Pi$ is calculated from the QP model.    

%%%%%%%%%%%%%%%%%%%%% Fig. 5 %%%%%%%%%%%%%%%%%%%%%%%%%%%%%%
\begin{figure*}[t]
\centering
\includegraphics[]{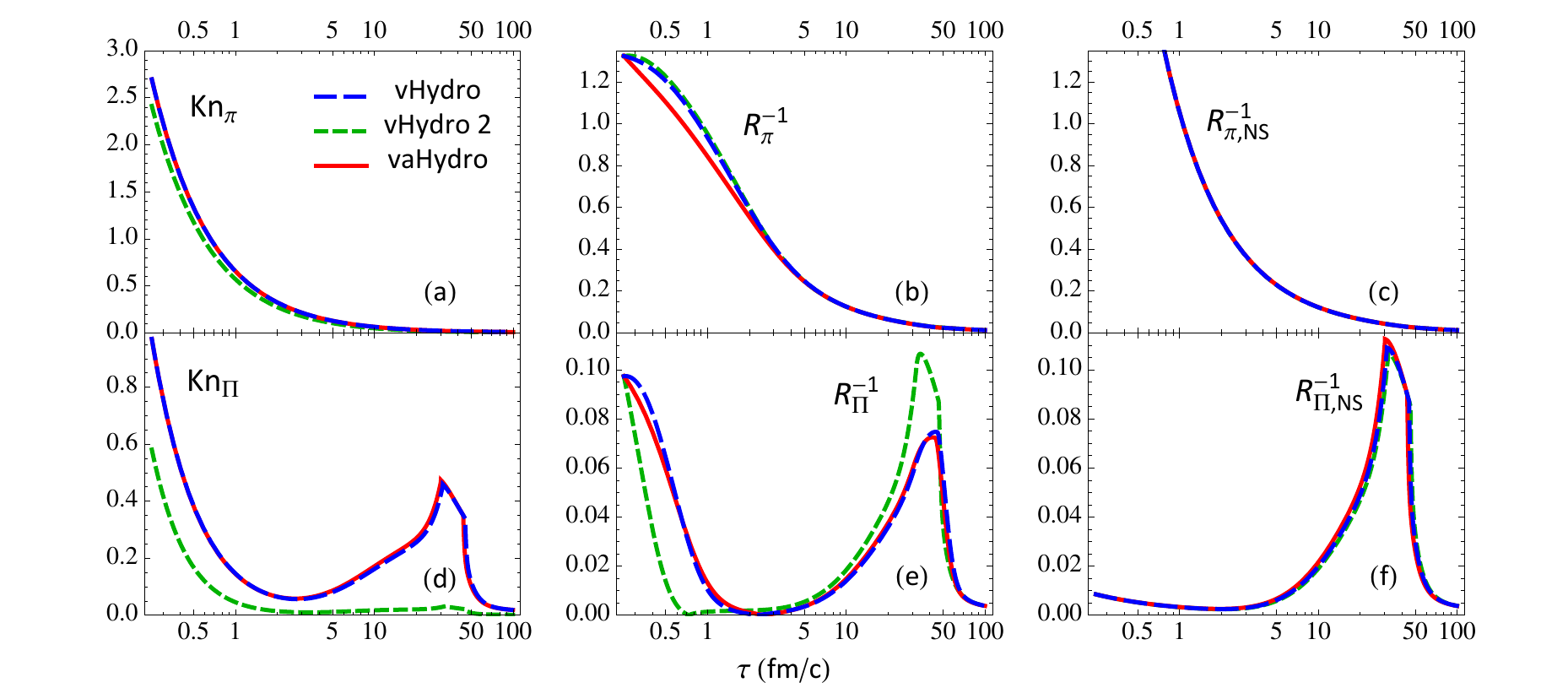}
\caption{(Color online)
	The shear and bulk Knudsen numbers and inverse Reynolds numbers associated with Fig.~\ref{F4}.
\label{F5}
}
\end{figure*}
%%%%%%%%%%%%%%%%%%%%%%%%%%%%%%%%%%%%%%%%%%%%%%%%%%%%%%

We have studied thermal equilibrium initial conditions with several other combinations of initial temperature $T_0$ and $\tau_0$, resulting in significantly different evolutions of the energy density and viscous pressure components (not shown here). Two features appear to be universal, however: (i) As long as we use transport coefficients computed from the same microscopic QP kinetic theory, the evolution of all components of the energy-momentum tensor, as well as of the mean field $B$, shows only very small differences (of the same order as shown in Fig.~\ref{F2}) between anisotropic and standard viscous hydrodynamics. (ii) Using instead transport coefficients for a Boltzmann gas of light fixed-mass particles, standard viscous hydrodynamics leads to significantly different evolutions for the bulk viscous pressure $\Pi$. For a meaningful comparison between anisotropic and standard viscous hydrodynamics it is therefore important that a consistent set of transport coefficients is being employed. Also, for a medium with broken conformal invariance (such as the quark-gluon plasma and other forms of QCD matter) non-conformal effects on the transport coefficients can have a large effect on the evolution of the bulk viscous pressure which may not be properly captured when using transport coefficients derived from a theory with weakly interacting degrees of freedom that have small masses. 

\vspace*{-3mm}
%%%%%%%%%%%%%%%%%%%%%%%%%%%%%%%%%%%%%%%%%%%%%%%%%%%%%%%
\subsection{Glasma-like initial conditions}
\label{sec5b}
%%%%%%%%%%%%%%%%%%%%%%%%%%%%%%%%%%%%%%%%%%%%%%%%%%%%%%%
\vspace*{-2mm}

In this subsection we repeat the exercise of the previous one for a different set of initial conditions, resembling those that one would get from matching the hydrodynamic evolution to a pre-equilibrium stage described by the IP-Glasma model \cite{Schenke:2012wb}. As already described, this model predicts approximately vanishing initial longitudinal pressure $\PL{\,\approx\,}0$ and $\Pperp{\,\approx\,}\ene/2$ \cite{Gelis:2013rba}. (In practice, we set $\PL/\Pperp = 0.01$ initially). We use the same initial longitudinal proper time and temperature as before. The corresponding results are plotted in Figs.~\ref{F4} and \ref{F5}.

%%%%%%%%%%%%%%%%%%% Fig. 6 %%%%%%%%%%%%%%%%%%%%%%%%%%%%%%%%%
%%%%%%%%%%%%%%%%%%%%%%%%%%%%%%%%%%%%%%%%%%%%%%%%%%%%%%%

For this extreme initial condition, the default magnitude of $B$ must be reduced by about $85\%$ in order to be able to successfully initialize the anisotropic microscopic parameters; for $B$ this is shown in Fig.~\ref{F4}c while the implications for the anisotropic microscopic parameters will be discussed in the following subsection. The highly non-equilibrium initial conditions manifest themselves in large starting values for the shear and bulk stresses and the non-equilibrium mean field. The initial shear stress (Fig.~\ref{F4}b) is about five times larger than its peak value for equilibrium initial conditions. The bulk viscous pressure (Fig.~\ref{F4}d) and non-equilibrium part of the mean field (Fig.~\ref{F4}f) are for the first fm/$c$ one to two orders of magnitude larger than for equilibrium initial conditions. In spite of this, anisotropic and standard viscous hydrodynamics still lead to almost identical evolution trajectories for the energy density (Fig.~\ref{F4}a) and viscous pressures (Figs.~\ref{F4}b,d) if QP transport coefficients are used, and if the latter are swapped out for those from a light Boltzmann gas, a significant change in the standard viscous hydrodynamic evolution is only seen for the bulk viscous pressure (short-dashed green curve in Fig.~\ref{F4}d). The shear stress $\pi$ and the pressure ratio $\PL/\Pperp$ emphasize the differences in the hydrodynamic models somewhat at early times (Figs.~\ref{F4}b,e), pushing the pressure ratio towards isotropy somewhat faster in anisotropic than in standard viscous hydrodynamics, but all three dynamical approaches converge to a common late-time behavior for $\pi$ and $\PL/\Pperp$ after about 2\,fm/$c$ (i.e. after about 3 times the initial shear relaxation time of about 0.8\,fm/$c$). It is, however, not the case that equilibrium and Glasma-like initial conditions lead to the same temperature evolution of the system: A careful comparison of Figs.~\ref{F2}a and \ref{F4}a shows that for the non-equilibrium initial conditions viscous heating by the large initial bulk and shear stresses causes the energy density (and therefore temperature) to drop somewhat more slowly than for equilibrium initial conditions, especially at early times. 

Figure~\ref{F4}f shows again that the ``asymptotic approximation'' (\ref{eq73}) for $\delta B^{(\mathrm{asy})}$ (dashed purple line) is not a good approximation for the full numerical evolution of $\delta B$ shown by the solid red line. Since for the Glasma-like initial conditions the non-equilibrium mean field contribution $\delta B$ is initially of the same order of magnitude as the equilibrium contribution $B_\eq$, the breakdown of this approximation is visible even in the evolution of the total mean field $B$ (solid red line) which is not at all described by $B^{(\mathrm{asy})}\equiv B_\eq{+}\delta B^{(\mathrm{asy})}$. 

Looking at the Knudsen and inverse Reynolds numbers in Fig.~\ref{F5} the only striking (although obvious) difference are the large starting values for both shear and bulk inverse Reynolds numbers when using Glasma-like initial conditions. Similar to the shear stress $\pi$ and pressure ratio $\PL/\Pperp$ in Figs.~\ref{F4}b,e, these two observables exhibit noticeable differences at early times between anisotropic and standard viscous hydrodynamic evolution.

%%%%%%%%%%%%%%%%%%%%%%%%%%%%%%%%%%%%%%%%%%%%%%%%%%%%%%%
\subsection{Evolution of the microscopic kinetic parameters}
\label{sec5b}
%%%%%%%%%%%%%%%%%%%%%%%%%%%%%%%%%%%%%%%%%%%%%%%%%%%%%%%

Although the parameters $(\Lambda,\alpha_\perp,\alpha_L)$ describing the slope and anisotropy of the momentum distribution of the microscopic degrees of freedom are vestiges from an underlying kinetic theory whose traces we have tried to erase as much as possible in our formulation of anisotropic hydrodynamics (hoping that eventually we can obtain the transport coefficients of QCD matter from a more fundamental approach), it is interesting to ``look under the hood'' and see how our parametrized anisotropic EOS works, i.e. how the QP model adjusts its microscopic parameters to accommodate the macroscopic anisotropic hydrodynamic initial conditions provided, and how it evolves them in response to the anisotropic hydrodynamic evolution of the energy-momentum tensor.

Figs.~\ref{F6}a,b compare, for equilibrium initial conditions, the evolution of the effective temperature parameter $\Lambda$ with that of the true temperature $T$ extracted from the energy density, and of the momentum anisotropy parameter $\alpha_L$ with that of $\alpha_\perp$, respectively. A comparison of Figs.~\ref{F6}a,b with Figs.~\ref{F3}b,e shows that large inverse Reynolds numbers in both the shear and bulk sectors correlate with effective temperatures $\Lambda{\,>\,}T$ and longitudinal momentum deformation parameter $\alpha_L{\,<\,}1$. Large shear inverse Reynolds numbers correlate with $\alpha_\perp$ deviating from unity in the opposite direction (i.e. with $\alpha_\perp{\,>\,}1$), leading to narrower longitudinal and wider transverse momentum distributions than in the equilibrium distribution $f_\eq$, consistent with $\PL/\Pperp<1$. Large bulk inverse Reynolds numbers push down both $\alpha_L$ and $\alpha_\perp$, corresponding to negative bulk viscous pressures. At late times, when both the shear and bulk inverse Reynolds numbers approach zero, the momentum distribution approaches local equilibrium, $\alpha_{L,\perp}\to1$ and $\Lambda\to T$.

For Glasma-like initial conditions, shown in Figs.~\ref{F6}c,d, these generic statements for the deformation parameters $\bm{\alpha}$ remain true but at early times the relationship between $T$ and $\Lambda$ is completely changed: the effective temperature $\Lambda$ starts out much smaller than the true temperature. A low effective temperature $\Lambda$ would narrow the microscopic momentum distribution in the transverse plane if it were not compensated by a very large ($\order(10)$) initial value of $\alpha_\perp$, which upholds the kinetic energy density and transverse pressure. On the other hand, $\alpha_L$ starts out almost at zero, reinforcing the narrowing of the longitudinal momentum distribution generated already by the small $\Lambda$ value and thereby causing a very small ratio of the kinetic contributions to $\PL$ and $\Pperp$. This is, of course, forced upon the system by the very anisotropic initial condition $\PL/\Pperp=0.01$. 
    
While the microscopic kinetic parameters $(\Lambda,\bm{\alpha})$ control only the kinetic contributions to energy density and pressures, the qualitative agreement of their tendencies extracted from this analysis of Figs.~\ref{F6}c,d with those of the total energy density and pressures shown in Fig.~\ref{F4} demonstrates that the mean field $B$, even where large, cannot alter the sign of the pressure anisotropy (shear stress). Its value shifts the average kinetic pressure relative to the kinetic energy density and thereby has a large influence on the bulk viscous pressure.    

%%%%%%%%%%%%%%%%%%%%%%%%%%%%%%%%%%%%%%%%%%%%%%%%%%%%%%%%%
\section{Conclusions and Outlook}
\label{sec6}
%%%%%%%%%%%%%%%%%%%%%%%%%%%%%%%%%%%%%%%%%%%%%%%%%%%%%%%%%
 
In this work we presented a purely macroscopic formulation of anisotropic hydrodynamics in 3+1 space-time dimensions, parametrized with Milne coordinates. To obtain the Lorentz structure of the anisotropic hydrodynamic equations, including the relaxation equations for the dissipative flows, we started from a microscopic description in terms of a relativistic Boltzmann-Vlasov equation with a relaxation-time approximated collision term. The mean field in the Boltzmann-Vlasov equation is constructed such that the energy density and equilibrium pressure of this kinetic theory satisfy an equation of state that agrees with the lattice QCD EOS of strongly interacting matter. The macroscopic equations of motion are derived from an anisotropic moment expansion of this Boltzmann-Vlasov equation, where the distribution function is split into a momentum-anisotropic leading order term $f_a$ and a residual correction $\dft$. To close the anisotropic moment expansion we use for the residual correction $\dft$ the 14-moment approximation. The leading-order term is constructed such that it can non-perturbatively account for the two largest dissipative effects encountered in relativistic heavy-ion collisions, a large longitudinal-transverse pressure anisotropy at early times and a large bulk viscous pressure during the phase transition of the matter from a quark-gluon plasma to color-confined hadronic matter. This requires the introduction of two momentum-anisotropy parameters $\alpha_L,\,\alpha_\perp$ into $f_a$ whose dynamics is fixed by a novel generalization of the Landau matching conditions that ensures that there are no residual corrections from $\dft$ to the longitudinal and transverse pressures of the system. This matching scheme allows us to completely eliminate the microscopic parameters that define $f_a$, and to write down, for the first time, a set of macroscopic anisotropic hydrodynamic evolution equations which make no explicit reference at all to their microscopic kinetic origin.

There are ten evolution equations for the ten components of the energy-momentum tensor. No specific assumptions are made for the equation of state relating the energy density and thermal pressure in thermal equilibrium, i.e. the equations can be used to describe any form of matter that behaves like a dissipative fluid. Two of these equations evolve the longitudinal and transverse pressures $\PL$ and $\Pperp$. Instead of splitting them into a thermal equilibrium pressure, a bulk viscous pressure and a longitudinal-transverse shear stress, with the latter two quantities assumed to be small and perturbatively treatable, in our approach $\PL$ and $\Pperp$ themselves are evolved, with the transport coefficients controlling how far they may deviate from the thermal equilibrium pressure.

The evolution of the energy-momentum tensor components is controlled by a standard set of driving forces, such as the longitudinal and transverse expansion rates, the various components of the velocity shear tensor, the flow vorticity, etc. In addition, the dissipative flows are characterized by a set of relaxation times describing their relaxation towards their first-order Navier-Stokes limits. Consistent with the anisotropic parametrization of the leading-order distribution $f_a$, the dissipative forces are separated into longitudinal and transverse parts using a systematic procedure involving orthogonal projection operators that was developed in Ref.~\cite{Molnar:2016vvu}. They are multiplied by a set of two dozen transport coefficients. These transport coefficients, as well as the relaxation times, are material properties of the dissipative fluid to be described.        

We do not know how to compute these transport coefficients for QCD matter from first principles. Therefore we use in this work a kinetic theory for weakly-interacting quasiparticles with temperature-dependent masses as a model for computing them. We write the distribution function for these quasiparticles as $f=f_a{+}\dft$ and parametrize $f_a$ in the same way as in the kinetic theory from which we first started. From the solution of the anisotropic hydrodynamic equations we then take the energy density $\ene$, longitudinal pressure $\PL$ and transverse pressure $\Pperp$, as well as the mean field $B$, and describe {\em the deviations} of $\PL$ and $\Pperp$ from the equilibrium pressure $\Peq(\ene)$ (which is taken from lattice QCD) in terms of the microscopic anisotropic parameters of the kinetic model. Having thus fixed the parameters of the kinetic model from the macroscopic hydrodynamic output, we can use it to compute all the transport coefficients in kinetic theory.

For the relaxation times we take previously introduced phenomenological parametrizations that were recently calibrated by a global comparison of a sophisticated dynamical model involving dissipative relativistic fluid dynamics at its core with experimental heavy-ion collision data collected at the LHC \cite{Bernhard:2016tnd,Bernhard:2017vql}.

As a first application of this new approach we have here studied the Bjorken evolution of a longitudinally boost-invariant, transversely homogeneous system, evolving it both with anisotropic and standard viscous hydrodynamics for comparison. We found remarkable agreement between the two approaches if both used quasiparticle transport coefficients but noticeable disagreements with a standard viscous hydrodynamic simulation using transport coefficients for a weakly interacting Boltzmann gas in the small fixed-mass limit. This suggests an unexpected robustness of the standard viscous hydrodynamic approach even in the presence of large shear and bulk viscous effects. A final assessment of the relative strengths and weaknesses of anisotropic vs. standard viscous hydrodynamics will, however, have to await the availability of full (3+1)-dimensional numerical evolution comparisons which are presently being pursued.

A key motivation for anisotropic hydrodynamics is that by accounting for the large dissipative components already at leading order, by parametrizing them into the leading-order distribution function $f_a$, the remaining dissipative flows arising from the residual deviation $\dft$ in the decomposition $f=f_a{+}\dft$ should be smaller than the dissipative flows in standard viscous hydrodynamics where they arise from $\delta f$ in the decomposition $f=f_\eq{+}\delta f$. Bjorken flow does not allow to test this expectation because for systems with Bjorken symmetry the residual dissipative flows arising from $\dft$ vanish anyhow exactly by symmetry. Full (3+1)-dimensional simulations will allow to answer this question. Taking the results reported in the last chapter of Ref.~\cite{Bazow:phdthesis} (based on a version of the present framework that did not treat the bulk viscous pressure non-perturbatively) as guidance for what to expect, anisotropic hydrodynamics as formulated here should indeed make the residual shear stress components significantly smaller in the center of the fireball, where the largest shear stresses are generated at early times by longitudinal expansion. However, the same may not necessarily hold for cells near the transverse edge of the fireball where the transverse expansion rate can exceed the longitudinal one and where accounting non-perturbatively for large effects associated with anisotropies {\em relative to the beam axis} (as we do here) may not offer significant advantages. We hope to be able to soon present numerical results that show how these expectations bear out in practice.

We close by noting that the observed sensitivity of the Bjorken evolution to the chosen model for computing the transport coefficients puts some urgency to the question how to compute the transport coefficients of anisotropic hydrodynamics from first principles for a theory such as hot and dense QCD. We have to leave this as a challenge for future work.     

%%%%%%%%%%%%%%%%%%%%%%%%%%%%%%%%%%%%%%%%%%%%%%%%%%%%%%%%
\acknowledgments

The authors would like to express their thanks to Lipei Du, Derek Everett, Mauricio Martinez, Etele Moln\'ar, Dirk Rischke, Chun Shen, Mike Strickland, Leonardo Tinti, and Gojko Vujanovic for fruitful discussions. This work was supported in part by the U.S. Department of Energy (DOE), Office of Science, Office for Nuclear Physics under Award No.\,\rm{DE-SC0004286} and within the framework of the BEST and JET Collaborations. The EOS Working Group of the BEST Collaboration, especially Paolo Parotto, is gratefully acknowledged for providing tables of the lattice QCD + hadron resonance gas equation of state used in this work. D.B., M.M. and U.H. acknowledge partial support by the National Science Foundation (NSF) within the framework of the JETSCAPE Collaboration under Award No. ACI-1550223. U.H.'s research was supported in part by the ExtreMe Matter Institute EMMI at the GSI Helmholtzzentrum f\"ur Schwerionenforschung, Darmstadt, Germany. He would also like to thank the Institut f\"ur Theoretische Physik of the J. W. Goethe-Universit\"at, Frankfurt, for kind hospitality.

%%%%%%%%%%%%%%%%%%%%%%%%%%%%%%%%%%%%%%%%%%%%%%%%%%%%%%%%

\appendix
%%%%%%%%%%%%%%%%%%%%%%%%%%%%%%%%%%%%%%%%%%%%%%%%%%%%%%%%
\section{Anisotropic integrals}
\label{appa}
%%%%%%%%%%%%%%%%%%%%%%%%%%%%%%%%%%%%%%%%%%%%%%%%%%%%%%%%
In this section, we define the anisotropic integrals that appear in this paper and show how to integrate them numerically. The anisotropic integrals $\I_{nrqs}$ and $\J_{nrqs}$ are defined as
\be
\label{eqA1}
   \I_{nrqs} =  \int_p \frac{(\up)^{n-r-2q}}{(2q)!!} \mzp^r \pxp^q (p \cdot \Omega \cdot p)^{s/2} f_a ,
\ee
\be
\label{eqA2}
   \J_{nrqs} = \int_p \frac{(\up)^{n-r-2q}}{(2q)!!} \mzp^r \pxp^q (p \cdot \Omega \cdot p)^{s/2} f_a \bar{f}_a,
\ee
where the distribution function $f_a$ is 
\be
\label{eqA3}
   f_a = f_{\eq}\left(\frac{1}{\Lambda} \sqrt{\frac{p^2_{\perp,\mathrm{LRF}}}{\alpha_\perp^2} + \frac{p^2_{z,\mathrm{LRF}}}{\alpha_L^2} + m^2}\,\right).
\ee
For particles with Boltzmann statistics $\I_{nrqs}=\J_{nrqs}$. Although there is no known analytical solution for these integrals for massive particles, their dimensionality can be reduced to one. After substituting the spherical coordinates
\be
\label{eqA4}
\begin{aligned}
   & p_{x,\mathrm{LRF}} = \alpha_\perp \Lambda \, \bar{p} \sin\theta \cos\phi, \\
   & p_{y,\mathrm{LRF}} = \alpha_\perp \Lambda \, \bar{p} \sin\theta \sin\phi, \\ 
   & p_{z,\mathrm{LRF}} = \alpha_L \Lambda \, \bar{p} \cos\theta,
\end{aligned}
\ee
with $\bar{p} = p / \Lambda$, the angular integrals in Eqs.~(\ref{eqA1},\ref{eqA2}) can be evaluated analytically, yielding 
\be
\label{eqA5}
\begin{split}
    \I_{nrqs} = & \, \frac{\alpha_\perp^{2q+2}\alpha_L^{r+1}\Lambda^{n+s+2}}{4\pi^2(2q)!!} 
                         \int_0^\infty d\bar p \, \bar p^{n+s+1} \times \\
                      & \R_{nrq}(\alpha_\perp, \alpha_L; \bar m / \bar p) f_{\eq}(\sqrt{\bar p^2 + \bar m^2}),
\end{split}
\ee
\be
\label{eqA6}
   \J_{nrqs} = \frac{\alpha_\perp^{2q+2}\alpha_L^{r+1}\Lambda^{n+s+2}}{4\pi^2(2q)!!}
   \int_0^\infty d\bar p \, \bar p^{n+s+1} \R_{nrq} \, f_{\eq} \, \bar{f}_{\eq},
\ee
where $\bar m = m / \Lambda$ and the functions $\R_{nrq}$ are defined as
\be
\label{eqA7}
\begin{split}
   & \R_{nrq}(\alpha_\perp, \alpha_L; \bar m / \bar p) =  w^{\,n-r-2q-1} \times \\
   & \int_{-1}^1 d \cos\theta \sin^{2q}\theta \cos^r\theta  (1 + z \sin^2\theta)^{(n-r-2q-1)/2},
\end{split}
\ee
with $w = \sqrt{\alpha_L^2 + (\bar m / \bar p)^2}$ and $z = \dfrac{\alpha_\perp^2 - \alpha_L^2}{w^2}$. The radial momentum integral can be computed numerically with generalized Gauss-Laguerre quadrature. For reference, we list the functions $\R_{nrq}$ that are needed in this paper:
\bs
\label{eqA8}
\beal
\R_{200} & = w \big(1+(1+z)t(z)\big) \\
\R_{220} & = \frac{-1+(1+z)t(z)}{z w} \\
\R_{201} & = \frac{1+(z-1)t(z)}{z w} \\
\R_{240} & = \frac{3+2z-3(1+z)t(z)}{z^2w^3} \\
\R_{202} & = \frac{3+z+(1+z)(z-3)t(z)}{z^2(1+z)w^3} \\
\R_{221} & = \frac{-3+(3+z)t(z)}{z^2w^3} \\
\R_{441} & = \frac{-15+13z+3(1+z)(5+z)t(z)}{4z^3w^3} \\
\R_{402} & = \frac{3(z-1)+(z(3z-2)+3)t(z)}{4z^2w} \\
\R_{421} & = \frac{3+z+(1+z)(z-3)t(z)}{4z^2w} \\
\R_{422} & = \frac{15+z+(z(z-6)-15)t(z)}{4z^3w^3} \\
\R_{403} & = \frac{(z-3)(5+3z)+3(1+z)(z(z-2)+5)t(z)}{4z^3(1+z)w^3}
\end{align}
\es 
where $t(z) = \dfrac{\arctan{\sqrt{z}}}{\sqrt{z}}$.
%

%%%%%%%%%%%%%%%%%%%%%%%%%%%%%%%%%%%%%%%%%%%%%%%%%%%%%%%%%
\section{Anisotropic integral identities}
\label{appe}
%%%%%%%%%%%%%%%%%%%%%%%%%%%%%%%%%%%%%%%%%%%%%%%%%%%%%%%%%

Here we show how to derive the identities (\ref{eq71}). First, we express the anisotropic integrals (\ref{eqA1}) and (\ref{eqA2}) as
\be
\label{eq:Iint}
   \I_{nrqs} = \frac{1}{(2q)!!} \int_p E_\mathrm{LRF}^{n-r-2q} \, p_{z,\mathrm{LRF}}^r \, p_{\perp,\mathrm{LRF}}^{2q} \, E_a^s \, f_a,
\ee
\be
\label{eq:Jint}
   \J_{nrqs} = \frac{1}{(2q)!!} \int_p E_\mathrm{LRF}^{n-r-2q} \, p_{z,\mathrm{LRF}}^r \, p_{\perp,\mathrm{LRF}}^{2q} \, E_a^s \, f_a \bar{f}_a,
\ee
where $E_a = \sqrt{m^2 + \dfrac{p_{\perp,\mathrm{LRF}}^2}{\alpha_\perp^2}+\dfrac{p_{z,\mathrm{LRF}}^2}{\alpha_L^2}}$. From here on out, we will suppress the LRF subscripts. To obtain the first identity (\ref{eq71}a), we introduce rapidity coordinates
\be
\label{eq:rapidity}
\begin{aligned}
   E= m_\perp \cosh{y}, \quad
%   p_x = p_x ,\
%   p_y = p_y, \
   p_z = m_\perp \sinh{y} \,,
\end{aligned}
\ee
where $m_\perp = \sqrt{m^2 + p_\perp^2}$, to rewrite~\eqref{eq:Jint} as
\be
\label{eq:Jint1}
\begin{split}
	\J_{nrqs} = \, & g \int \frac{dy \,d^2 p_\perp}{(2\pi)^3(2q)!!} \, m_\perp^{n-2q} (\cosh{y})^{n-r-2q} \\
			     & \times \,(\sinh{y})^r \, p_\perp^{2q} \, E_a^s \, f_a \bar{f}_a .
\end{split}
\ee
Next, one can write the distribution term $f_a \bar{f}_a$ as
\be
\label{eq:ffbar_rapidity}
	f_a \bar{f}_a = - \frac{\Lambda \, \alpha_L^2 \, E_a}{m_\perp^2 \cosh{y} \, \sinh{y}} \frac{\partial f_a}{\partial y} .
\ee
After integrating by parts with respect to the variable $y$ (where the boundary term vanishes for $r \geq 2$) one obtains the relation 
\begin{eqnarray}
\label{eq:id1} 
    \J_{nrqs} &=&\Lambda \, \alpha_L^2 \,(n{-}r{-}2q{-}1)\,\I_{n-2,r,q,s+1} 
\\\nonumber
    &&+ \Lambda \, \alpha_L^2 \,(r{-}1)\,\I_{n-2,r-2,q,s+1} + \Lambda\,(s{+}1)\,\I_{n,r,q,s-1} \,,
\end{eqnarray}
which for $(n,r,q,s) = (4,2,0,-1)$ yields Eq.~(\ref{eq71}a).

For the second identity (\ref{eq71}b) one uses cylindrical coordinates
\be
\label{eq:cylindrical}
\begin{aligned}
    E = \sqrt{m^2{+}p_\perp^2{+}p_z^2},  \ 
    p_x = p_\perp \cos\phi, \ 
    p_y = p_\perp \sin\phi \ 
%    p_z = p_z \,,
\end{aligned}
\ee
to express the integral~\eqref{eq:Jint} as
\be
\label{eq:Jint2}
\begin{split}
	\J_{nrqs} = \, & g \int \frac{dp_\perp dp_z}{(2\pi)^2(2q)!!} \, E^{n-r-2q-1} \,
	p_z^{r} \, p_\perp^{2q+1} \, E_a^s \, f_a \bar{f}_a .
\nonumber
\end{split}
\ee
The term $f_a \bar{f}_a$ can be written as
\be
\label{eq:ffbar_cylinder}
	f_a \bar{f}_a = - \frac{\Lambda \, \alpha_\perp^2 \, E_a}{p_\perp} \frac{\partial f_a}{\partial p_\perp} .
\ee
Integration by parts with respect to $p_\perp$, with the boundary term vanishing for $q \geq 1$, gives
\be
\begin{split}
\label{eq:id2} 
\J_{nrqs} &= \Lambda \, \alpha_\perp^2 \,(n{-}r{-}2q{-}1)\,\I_{n-2,r,q,s+1}  \\
	       &+\Lambda \, \alpha_\perp^2 \,\I_{n-2,r,q-1,s+1} + \Lambda\,(s{+}1)\,\I_{n,r,q,s-1} \,,
\end{split}
\ee
which for $(n,r,q,s) = (4,0,1,-1)$ yields Eq.~(\ref{eq71}b). 
 
Finally, to get the third identity (\ref{eq71}c) we use spherical coordinates
\be
\label{eq:spherical}
\begin{aligned}
   &E = \sqrt{m^2 + p^2},  \quad p_z = p \cos\theta \,,\\
   &p_x = p \sin\theta \cos\phi, \  p_y = p \sin\theta \sin\phi \ 
\end{aligned}
\ee
in the integral~\eqref{eq:Jint}
\be
\label{eq:Jint3}
\begin{split}
	\J_{nrqs} = \, & g \int \frac{dp \, d(\cos\theta)}{(2\pi)^2(2q)!!} \, E^{n-r-2q-1} \, p^{r+2q+2}  \\
			     & \times \, (\cos\theta)^r \, (\sin\theta)^{2q} \, E_a^s \, f_a \bar{f}_a 
\end{split}
\ee
and write the term $f_a \bar{f}_a$ as
\be
\label{eq:ffbar_spherical}
    f_a \bar{f}_a = - \frac{\Lambda \, \alpha_\perp^2 \alpha_L^2}{\alpha_\perp^2{-}\alpha_L^2} \, 
                              \frac{E_a}{p^2\cos\theta}\, \frac{\partial f_a}{\partial \cos\theta} .
\ee
Integrating by parts with respect to $\cos\theta$, where the boundary term vanishes for $q \geq 1$, gives the following relation:
\begin{eqnarray}
\label{eq:id3} 
   \J_{nrqs} &=& \frac{\Lambda \, \alpha_\perp^2 \alpha_L^2}{\alpha_\perp^2{-}\alpha_L^2} 
                          \Bigl((r{-}1)\,\I_{n-2,r-2,q,s+1}{-}\I_{n-2,r,q-1,s+1}\Bigr) 
\nonumber\\
	            &+& \Lambda\,(s{+}1)\,\I_{n,r,q,s-1} \,,
\end{eqnarray}
which for $(n,r,q,s) = (4,2,1,-1)$ yields Eq.~(\ref{eq71}c).
 
%%%%%%%%%%%%%%%%%%%%%%%%%%%%%%%%%%%%%%%%%%%%%%%%%%%%%%%%%
\section{Anisotropic transport coefficients}
\label{appb}
%%%%%%%%%%%%%%%%%%%%%%%%%%%%%%%%%%%%%%%%%%%%%%%%%%%%%%%%%

We list the transport coefficients that appear in the relaxation equations \eqref{eq33}--\eqref{eq36}. Some of the expressions contain terms $\propto dm/dT$ (see footnote~\ref{fn5}). For the quasiparticle model described in Sec~\ref{sec4a}, the temperature-dependent mass $m(T)$ and its derivative $dm/dT$ should be taken from Fig.~\ref{F1}.

The coefficients controlling the evolution of the kinetic longitudinal pressure $\PL^{(k)}$ are 
\bs
\label{eqB1}
\beal
\bar{\zeta}^{L(k)}_z & = \I_{2400} - 3 \PL^{(k)} + m\frac{dm}{dT}\frac{dT}{d\ene}(\ene{+}\PL)\I_{0200},  \\
\bar{\zeta}^{L(k)}_\perp & = \I_{2210} - \PL^{(k)} + m\frac{dm}{dT}\frac{dT}{d\ene}(\ene{+}\Pperp)\I_{0200}, \\
\bar{\lambda}^{L(k)}_{Wu} & = \frac{\J_{4410}}{\J_{4210}} + m\frac{dm}{dT}\frac{dT}{d\ene}\I_{0200},\\
\bar{\lambda}^{L(k)}_{W\perp} & =  1- \bar{\lambda}^{L(k)}_{Wu},\\
\bar{\lambda}^{L(k)}_\pi & = \frac{\J_{4220}}{\J_{4020}} + m\frac{dm}{dT}\frac{dT}{d\ene}\I_{0200}. 
\end{align}
\es
Those controlling the evolution of the kinetic transverse pressure $\Pperp^{(k)}$ are
\bs
\label{eqB2}
\beal
\bar{\zeta}^{\perp(k)}_z & = \I_{2210} - \Pperp^{(k)} + m\frac{dm}{dT}\frac{dT}{d\ene}(\ene{+}\PL)\I_{0010}, \\
\bar{\zeta}^{\perp(k)}_\perp & = 2\bigl(\I_{2020} - \Pperp^{(k)} \bigr) 
                                         + m\frac{dm}{dT}\frac{dT}{d\ene}(\ene{+}\Pperp)\I_{0010}, \\
\bar{\lambda}^{\perp(k)}_{W\perp} & = \frac{2 \,\J_{4220}}{\J_{4210}} + m\frac{dm}{dT}\frac{dT}{d\ene}\I_{0010},\\
\bar{\lambda}^{\perp(k)}_{Wu} & = \bar{\lambda}^{\perp(k)}_{W\perp} - 1, \\
\bar{\lambda}^{\perp(k)}_\pi & = 1 - \frac{3\,\J_{4030}}{\J_{4020}} -  m\frac{dm}{dT}\frac{dT}{d\ene}\I_{0010}.
\end{align}
\es
The evolution of the longitudinal momentum diffusion current $\Wperp$ involves the coefficients
\bs
\label{eqB3}
\beal
\bar{\eta}^W_u & = {\textstyle\frac{1}{2}}\bigl(\PL^{(k)}  - \I_{2210}\bigr), \\
\bar{\eta}^W_\perp & = {\textstyle\frac{1}{2}}\bigl(\Pperp^{(k)}  - \I_{2210}\bigr), \\
\bar{\tau}^W_z & = \PL^{(k)}  - \Pperp^{(k),}  \\
\bar{\delta}^W_{W} & = \bar{\lambda}^W_{W\perp}{-}{\textstyle\frac{1}{2}} 
                                   + m\frac{dm}{dT}\frac{dT}{d\ene}(\ene{+}\Pperp)\left(\frac{\J_{2210}}{\J_{4210}}\right), \\
\bar{\lambda}^W_{Wu} & = 2 -  \frac{\J_{4410}}{\J_{4210}} 
                                         - m\frac{dm}{dT}\frac{dT}{d\ene}(\ene{+}\PL)\left(\frac{\J_{2210}}{\J_{4210}}\right) ,\\
\bar{\lambda}^W_{W\perp} & = \frac{2 \,\J_{4220}}{\J_{4210}} - 1,\\
\bar{\lambda}^W_{\pi u} & = \frac{\J_{4220}}{\J_{4020}} \\
\bar{\lambda}^W_{\pi\perp} & = \bar{\lambda}^W_{\pi u}  - 1 , 
\end{align}
\es
while that of the transverse shear stress tensor $\piperp$ involves the coefficients
\bs
\label{eqB4}
\beal
\bar{\eta}_\perp & = \Pperp^{(k)}  - \I_{2020}, \\
\bar{\delta}^\pi_\pi & = {\textstyle\frac{3}{4}} \bar{\tau}^\pi_\pi + {\textstyle\frac{1}{2}}  
                                 - m\frac{dm}{dT}\frac{dT}{d\ene}(\ene{+}\Pperp)\left(\frac{\J_{2020}}{\J_{4020}}\right), \\
\bar{\tau}^\pi_\pi & =  2 - \frac{4\,\J_{4030}}{\J_{4020}} , \\
\bar{\lambda}^\pi_{\pi} & = \bar{\lambda}^W_{\pi u} - 1 
                                       + m\frac{dm}{dT}\frac{dT}{d\ene}(\ene{+}\PL)\left(\frac{\J_{2020}}{\J_{4020}}\right), \\
\bar{\lambda}^\pi_{Wu} & =  \bar{\lambda}^W_{W\perp} -  1,\\
\bar{\lambda}^\pi_{W\perp} & = \bar{\lambda}^\pi_{Wu} + 2.
\end{align}
\es
Additional transport coefficients appear for systems with conserved charges and associated diffusion currents \cite{Molnar:2016vvu}.

%%%%%%%%%%%%%%%%%%%%%%%%%%%%%%%%%%%%%%%%%%%%%%%%%%%%%%
\section{Transport coefficients for $\PL$ and $\Pperp$ in the quasiparticle model}
\label{appc}
%%%%%%%%%%%%%%%%%%%%%%%%%%%%%%%%%%%%%%%%%%%%%%%%%%%%%%

Here we redefine the transport coefficients controlling the macroscopic longitudinal and transverse pressures after adding to Eqs.~(\ref{eq33})-(\ref{eq34}) the mean field terms from Eq.~(\ref{eq52}). The transport coefficients for the total longitudinal pressure $\PL$ are
\bs
\label{eqC1}
\beal
   \bar{\zeta}^L_z & = \bar{\zeta}^{L(k)}_z 
   - \frac{1}{m}\frac{dm}{dT}\frac{dT}{d\ene}(\ene{+}\PL)\Bigl(\ene^{(k)}{-}2\Pperp^{(k)}{-}\PL^{(k)}\Bigr),
\\
   \bar{\zeta}^L_\perp & = \bar{\zeta}^{L(k)}_\perp 
   - \frac{1}{m}\frac{dm}{dT}\frac{dT}{d\ene}(\ene{+}\Pperp)\Bigl(\ene^{(k)}{-}2\Pperp^{(k)}{-}\PL^{(k)}\Bigr),
\\
   \bar{\lambda}^L_{Wu} & =\bar{\lambda}^{L(k)}_{Wu} 
   - \frac{1}{m}\frac{dm}{dT}\frac{dT}{d\ene}\Bigl(\ene^{(k)}{-}2\Pperp^{(k)}{-}\PL^{(k)}\Bigr),
\\
   \bar{\lambda}^L_{W\perp} & =  1 - \bar{\lambda}^L_{Wu},
\\
   \bar{\lambda}^L_\pi & = \bar{\lambda}^{L(k)}_\pi  
   + \frac{1}{m}\frac{dm}{dT}\frac{dT}{d\ene}\Bigl(\ene^{(k)}{-}2\Pperp^{(k)}{-}\PL^{(k)}\Bigr).
\end{align}
\es
Those controlling the evolution of the total transverse pressure $\Pperp$ are
\bs
\label{eqC2}
\beal
   \bar{\zeta}^\perp_z & = \bar{\zeta}^{\perp(k)}_z 
   - \frac{1}{m}\frac{dm}{dT}\frac{dT}{d\ene}(\ene{+}\PL)\Bigl(\ene^{(k)}{-}2\Pperp^{(k)}{-}\PL^{(k)}\Bigr),
\\
   \bar{\zeta}^\perp_\perp & = \bar{\zeta}^{\perp(k)}_\perp 
   - \frac{1}{m}\frac{dm}{dT}\frac{dT}{d\ene}(\ene{+}\Pperp)\Bigl(\ene^{(k)}{-}2\Pperp^{(k)}{-}\PL^{(k)}\Bigr),
\\
   \bar{\lambda}^\perp_{W\perp} & = \bar{\lambda}^{\perp(k)}_{W\perp} 
   - \frac{1}{m}\frac{dm}{dT}\frac{dT}{d\ene}\Bigl(\ene^{(k)}{-}2\Pperp^{(k)}{-}\PL^{(k)}\Bigr),
\\
   \bar{\lambda}^\perp_{Wu} & = \bar{\lambda}^\perp_{W\perp} - 1, 
\\
   \bar{\lambda}^\perp_\pi & = \bar{\lambda}^{\perp(k)}_\pi 
   + \frac{1}{m}\frac{dm}{dT}\frac{dT}{d\ene}\Bigl(\ene^{(k)}{-}2\Pperp^{(k)}{-}\PL^{(k)}\Bigr).
\end{align}
\es
%

%%%%%%%%%%%%%%%%%%%%%%%%%%%%%%%%%%%%%%%%%%%%%%%%%%%%%%
\section{Viscous hydrodynamic equations}
\label{appd}
%%%%%%%%%%%%%%%%%%%%%%%%%%%%%%%%%%%%%%%%%%%%%%%%%%%%%%

Here we derive the viscous hydrodynamic equations~\eqref{eq:vhydroeqs} and their transport coefficients. We start with the quasiparticle case (long-dashed blue) and derive the relaxation equation for $\delta B$ and its second-order solution \eqref{eq:dB2nd}. The general evolution equation for $\delta B$ is given by \cite{Tinti:2016bav}
\be
\delta\dot B = - \frac{\delta B}{\tau_\Pi} + \frac{\dot m}{m} \left(3 \Pi + 4 \,\delta B\right),
\ee
where the $\dot m / m$ term can be written as
\be
\frac{\dot m}{m} = \frac{1}{m} \frac{dm}{dT}\frac{dT}{d\ene} \dot\ene.
\ee
We replace the time derivative $\dot\ene$ with the energy conservation law in viscous hydrodynamics
\be
\label{eq:energyvhydro}
\dot\ene + (\ene + \Peq + \Pi)\theta - \pi^\munu \sigma_\munu = 0 \,,
\ee
where $\theta = \del_\mu u^\mu$ is the scalar expansion rate and $\sigma_\munu = \Delta^{\alpha\beta}_\munu \del_\beta u_\alpha$ is the velocity shear tensor. For the second-order relaxation equation, we only need the first-order approximation $\dot\ene \approx - (\ene + \Peq)\theta$, thus
\be
\frac{\dot m}{m} \approx - \frac{1}{m} \frac{dm}{dT}\frac{dT}{d\ene} (\ene + \Peq)\theta .
\ee
The equation of motion for $\delta B$ then reduces to
\be
\label{eq:dBdot1}
\delta\dot B = - \frac{\delta B}{\tau_\Pi} - \frac{\ene+\Peq}{m} \frac{dm}{dT}\frac{dT}{d\ene} \left(3 \Pi + 4 \,\delta B\right) \theta.
\ee
To first order in deviations from equilibrium $\delta B$ = 0. The second-order solution is given by \eqref{eq:dB2nd} \cite{Tinti:2016bav}. In Eq.~\eqref{eq:dBdot1}, we truncate the third-order term $\propto \delta B \, \theta$ to arrive at the second-order relaxation equation
\be
\label{eq:dBdot2}
\delta\dot B = - \frac{\delta B}{\tau_\Pi} - \frac{3(\ene+\Peq)}{m} \frac{dm}{dT}\frac{dT}{d\ene} \Pi\, \theta.
\ee

Next, we derive the relaxation equations for the viscous components $\Pi$ and $\pi^\munu$ in the same manner as in Sec. \ref{sec3}. We start by taking the time derivative of their quasi-kinetic definitions \cite{Tinti:2016bav}
\bs
\label{eq:vhydrodot1}
\beal
&\dot{\Pi} =  \frac{1}{3} \, D \int_p (- p \cdot \Delta \cdot p)  \delta f - \delta \dot B, \\
&\dot{\pi}^{\langle\munu\rangle} =  \frac{1}{3} \Delta^\munu_{\alpha\beta} \, D \int_p  p^{\langle\alpha} \, p^{\beta\rangle}  \delta f \,,
\end{align}
\es
where $\delta f$ is the non-equilibrium correction to the distribution function
\be
\label{eq:fvhydro}
f = f_\eq + \delta f \,,
\ee
with $f_\eq = \exp\big({-}u{\,\cdot\,}p/T\big)$ being the local equilibrium Boltzmann distribution. The time derivative $\delta \dot B$ is given by~\eqref{eq:dBdot2}. We substitute the terms containing $\delta\dot{f}$ using the Boltzmann equation~\eqref{eq13} and Eq.~\eqref{eq:fvhydro}:
\be
\begin{split}
\delta\dot{f} = \, &- \dot{f}_\eq + \frac{C[f] - m \, \del^\mu m \, \del_\mu^{(p)} f}{\up}  \\
&- \frac{p^{\langle\mu\rangle} \nabla_\mu f_\eq}{\up} - \frac{p^{\langle\mu\rangle} \nabla_\mu \delta f}{\up} \,,
\end{split}
\ee
where $p^{\langle\mu\rangle} = \Delta^\mu_\nu p^\nu$ and $\nabla_\mu = \Delta^\nu_\mu \del_\nu$ is the spatial gradient. To close the system of equations we use the 14-moment approximation for $\delta f$ \cite{Denicol:2014vaa}
\be
\label{eq:14ansatz}
   \frac{\delta f}{f_\eq} = c_\ene (\up)^2 + \frac{1}{3}c_\Pi \,(- p \cdot \Delta \cdot p) 
   + c_{\pi}^{\langle\munu\rangle} p_{\langle\mu} \, p_{\nu\rangle},
\ee
where $p_{\langle\mu} \, p_{\nu\rangle}$ = $\Delta^{\alpha\beta}_\munu p_\alpha p_\beta$. To solve for the coefficients we insert the expression~\eqref{eq:14ansatz} into the energy matching condition and the definitions of $\Pi$ and $\pi^\munu$:  
\bs
\label{eq:14momenteq}
\beal
&\delta\ene = \int_p (\up)^2 \delta f = 0, \\
&\Pi =  - \frac{1}{3} \int_p p_\mu p_\nu \Delta^\munu  \delta f, \\
&\pi^\munu = \int_p p_{\langle\mu} \, p_{\nu\rangle} \delta f.
\end{align}
\es
Since the 14-moment approximation is first-order in the dissipative flows, we neglect the second-order contribution ${\sim\,}\delta B$ to the energy density and bulk viscous pressure in~\eqref{eq:14momenteq}. After some algebra, the coefficients are
\bs
\beal
&c_\ene = - \frac{\I_{41} \, \Pi}{\frac{5}{3}\I_{40} I_{42}- \I_{41}^2}, \\
&c_\Pi = \frac{\I_{40} \, \Pi}{\frac{5}{3}\I_{40} I_{42}- \I_{41}^2}, \\
&c_\pi^{\langle\munu\rangle} = \frac{\pi^\munu}{2\,\I_{42}},
\end{align}
\es
where we defined the thermodynamic integrals
\be
\label{eq:intI}
  \I_{nq} =  \int_p \frac{(\up)^{n-2q}}{(2q+1)!!}(- p \cdot \Delta \cdot p)^q f_\eq.
\ee
The final expression for $\delta f$ is
\be
\label{eq:14momenteq2}
\frac{\delta f}{f_\eq} = \big(\bar{c}_\ene (\up)^2 + \frac{1}{3}\bar{c}_\Pi \,(- p \cdot \Delta \cdot p)\big)\Pi + \frac{1}{2}\bar{c}_{\pi} p_{\langle\mu} \, p_{\nu\rangle} \pi^\munu \,,
\ee
where $\bar{c}_\ene = c_\ene / \Pi$, $\bar{c}_\Pi = c_\Pi / \Pi$ and $\bar{c}_\pi = 1 / \I_{42}$. After integration by parts and inserting for $\delta f$ the 14-moment approximation~\eqref{eq:14momenteq2}, the relaxation equations for $\Pi$ and $\pi^\munu$ reduce to
\bs
\label{eq:relaxvhydro}
\beal
&\dot{\Pi} = -\frac{\Pi}{\tau_\Pi} - \beta_\Pi \theta - \delta_{\Pi\Pi} \Pi \theta + \lambda_{\Pi\pi}\pi^\munu \sigma_\munu,
\\
&\begin{aligned}
\dot{\pi}^{\langle\munu\rangle} = \, & -\frac{\pi^\munu}{\tau_\pi} - \beta_\pi \sigma^\munu + 2\pi^{\lambda\langle\mu} \omega^{\nu\rangle}_\lambda \\
& - \tau_{\pi\pi} \pi^{\lambda\langle\mu} \sigma^{\nu\rangle}_\lambda - \delta_{\pi\pi} \pi^\munu \theta + \lambda_{\pi\Pi} \Pi \sigma^\munu.
\end{aligned}
\end{align}
\es
Here $\omega^\munu = \Delta^\mu_\alpha \Delta^\nu_\beta \del_{[\beta} u_{\alpha ]}$ is the vorticity tensor. The transport coefficients are
\bs
\beal
&\beta_\pi = \frac{\I_{32}}{T} ,\\
&\beta_\Pi = \frac{5}{3}\beta_\pi - c_s^2(\ene+\Peq) + c_s^2 m \frac{dm}{dT} \, \I_{11}, \\
&\begin{aligned}
\delta_{\Pi\Pi} = \, & 1 - c_s^2 - \frac{m^4}{9}(\bar{c}_\ene \, \I_{00} + \bar{c}_\Pi \, \I_{01}) \\
& - m \frac{dm}{dT}\frac{dT}{d\ene}(\ene+\Peq)\Big(\bar{c}_\ene \, \I_{21} + \frac{5}{3} \bar{c}_\Pi \, \I_{22} + \frac{3}{m^2} \Big),
\end{aligned}  \\
&\lambda_{\Pi\pi} = \frac{1}{3} - c_s^2 + \frac{\bar{c}_\pi m^2 \I_{22}}{3}, \\
&\tau_{\pi\pi} = \frac{10}{7} + \frac{4 \,\bar{c}_\pi m^2 \I_{22}}{7}, \\
&\delta_{\pi\pi} = \frac{4}{3} + \frac{\bar{c}_\pi m^2\I_{22}}{3} - \bar{c}_\pi m \frac{dm}{dT}\frac{dT}{d\ene}(\ene+\Peq) \I_{22} ,\\
&\lambda_{\pi\Pi} = \frac{6}{5} - \frac{2 m^4}{15} (\bar{c}_\ene \, \I_{00} + \bar{c}_\Pi \, \I_{01}).
\end{align}
\es
In an expansion in powers of $z = m/T \ll 1$, taking the fixed-mass limit $dm/dT = 0$, the leading terms for these transport coefficients are \cite{Denicol:2014vaa}
\bs
\beal
&\beta_\pi = \frac{\ene_\eq^{(k)}+\Peq^{(k)}}{5} + \order(z^2), \\
&\beta_\Pi = 15\Big(\frac{1}{3} - \big(c^{(k)}_s\big)^2\Big)^2\big(\ene_\eq^{(k)}+\Peq^{(k)}\big) + \order(z^5),\\
&\delta_{\Pi\Pi} = \frac{2}{3} + \order(z^2 \ln z), \\
&\lambda_{\Pi\pi} = \frac{8}{5}\Big(\frac{1}{3} -  \big(c^{(k)}_s\big)^2\Big) + \order(z^4), \\
&\tau_{\pi\pi} = \frac{10}{7} + \order(z^2) ,\\
&\delta_{\pi\pi} = \frac{4}{3} + \order(z^2), \\
&\lambda_{\pi\Pi} = \frac{6}{5} + \order(z^2 \ln z).
\end{align}
\es
Here $\big(c^{(k)}_s\big)^2 = \I_{31}/\I_{30}$ is the kinetic theory definition for the squared speed of sound. Although $\ene_\eq^{(k)}$, $\Peq^{(k)}$ and $\big(c^{(k)}_s\big)^2$ are kinetic theory expressions, it is common practice to replace them in the above expressions with those from lattice QCD, and we will do so here. For Bjorken flow, the energy conservation law~\eqref{eq:energyvhydro} and relaxation equations~\eqref{eq:relaxvhydro} simplify greatly. The gradient terms are $\sigma^\munu = \diag(0,\frac{1}{3\tau}, \frac{1}{3\tau},-\frac{2}{3\tau^3})$, $\theta = 1 / \tau$, and $\omega^\munu = 0$. The shear stress components are $\pi^\munu = \diag(0, - \frac{1}{2} \tau^2 \pi^{\eta\eta},- \frac{1}{2} \tau^2 \pi^{\eta\eta},\pi^{\eta\eta})$. As a result, the viscous hydrodynamic equations reduce to \eqref{eq:vhydroeqs}.

%%%%%%%%%%%%% Bibliography %%%%%%%%%%%%%%%%%%%%%%%%

\bibliography{vahydro2}
  
%%%%%%%%%%%%%%%%%%%%%%%%%%%%%%%%%%%%%%%%%%%%  
    
\end{document}